\documentclass[11pt]{article}
\usepackage{amsmath,amssymb}
\usepackage[affil-it]{authblk}

\usepackage{units}
\usepackage{subfigure}
\usepackage{graphicx}

\usepackage{eurosym}

\usepackage{float}

\graphicspath{{Pics/}}

\newcommand{\be}{\begin{equation}}
\newcommand{\ee}{\end{equation}}
\newcommand{\bea}{\begin{eqnarray}}
\newcommand{\eea}{\end{eqnarray}}

\begin{document}

\title{
Sensitivity studies of heat transfer: forced convection across a cylindrical pipe and duct flow.
}

\author[1]{A. Ferrantelli
 \thanks{Electronic address: \texttt{andrea.ferrantelli@aalto.fi}; Corresponding author}}
 
 \author[2]{P. Mel\'ois}

  \author[1]{M. Viljanen}

\affil[1]{Aalto University, Department of Civil and Structural Engineering,
FI-00076 Aalto, Finland}
\affil[2]{
Institut National des Sciences Appliqu\'ees de Toulouse\\
 31077 Toulouse Cedex 4, France
}

\date{\today}


\maketitle

\setcounter{table}{0}

\begin{abstract}
	We consider two common heat transfer processes and perform a through sensitivity study of the variables involved.
	We derive and discuss analytical formulas for the heat transfer coefficient in function of film velocity, air temperature and pipe diameter. The according plots relate to a qualitative analysis of the multi-variable function $h$, according to functional optimization.
	For each process, we provide with graphs and tables of the parameters of interest, such as the Reynolds number. This method of study and the specific values can constitute a useful reference for didactic purposes.

\end{abstract}


\newpage


\section{Introduction}

Studies of heat transfer pertain fundamental physics which is crucial in many situations of practical interest, for instance in building physics. As such, this topic is included in most academic curricula for engineers. Here we address the subject from the perspective of building physics, a field of study that blends physics, mathematics and engineering. A through knowledge of heat transfer has indeed diverse applications which range from indoor environmental quality, moisture tolerance and overall energy saving \cite{Hens}.

	A significant improvement in this field took place after the energy crises in the 1970s, and since then the technical literature developed very quickly. Especially in the last decade, it offered a number of studies which benefit from new technology developments such as computer simulations (see for instance \cite{Review} and references quoted therein).

	The present paper addresses an approach to calculations which can effectively support the teaching in engineering as a whole, as it is not only restricted to heat transfer. We will show that examining the physics behind the specific process gives interesting insights on the phenomenology, especially on didactic but also on research grounds.
Here we focus on heat transfer processes, listing tables and plots of the quantities of interest\footnote{This method is similar to the so-called \emph{functional optimization}, which has been adapted to engineering studies in \cite{Opt}.}. 

The main tool here is a through study of the relevant formulas, focused on the structure of the equations. For instance, studying how the Reynolds number or the convection coefficient depend on the temperature through viscosity and density.

	This constitutes a novelty of this work. It is a common procedure in other fields, such as theoretical physics, by students and researchers. However, qualitative aspects are still quite overlooked in engineering in favor of numerics, due to the enormous development of simulation programs \cite{Review}. A through analytical study should anyway be considered, as it clearly complements and optimizes the numerical computations.
	
	With this approach, the student is indeed able to predict the behavior of the crucial quantities \emph{a priori}, and knows how these behave in function of the system parameters, even before starting the numerical calculations. This is useful for having more efficient simulations and to validate the according results first at the theoretical level.

	As an example, we apply this analytical method to two common heat transfer processes, namely to fluid flow across a cylindrical pipe in Section 2 and to internal flow in a duct in Section 3.
	We provide with

- interpolation curves of the physical properties of the fluids considered

- a qualitative analysis of the main parameters, namely the Reynolds number and the heat transfer coefficient

- a quantitative study with tables and graphs of the according values.
This is especially useful for students, who generally do \emph{not} have knowledge of the \emph{actual} numbers.
Moreover, all the values and the materials (fluids) have been chosen so that the results can be useful for teaching and also in practical situations.

	In the Appendix we show how to use the least square method to compute interpolation curves of the variables of interest. As a specific example, we consider the critical length for a flow on a flat plate, in the case of engine oil and dry air.


\section{Heat transfer across a cylindrical pipe}

\begin{figure}[ht]
\centering
\includegraphics[width=0.8\textwidth]{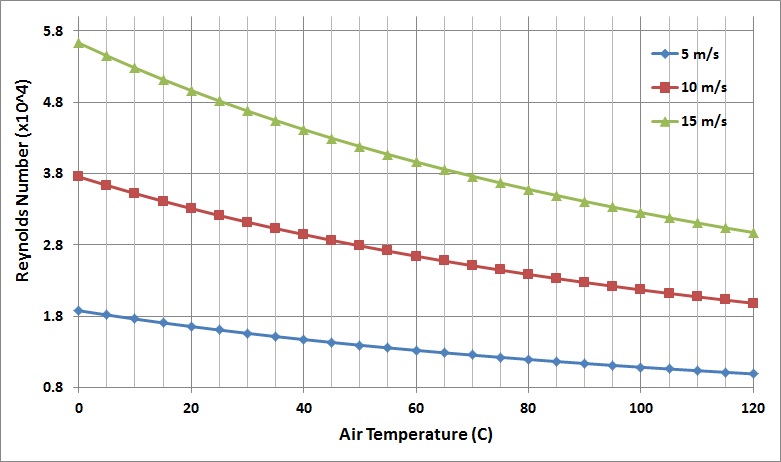}
\caption{${\rm Re}_D$ for air in function of $T$, for $D=\unit[5]{cm}$ and $V=5, 10, 15\,{m/s}$, as given by Eq.(\ref{Reynolds}).}
\label{fig:ReynT}
\vspace{3mm}
\includegraphics[width=0.8\textwidth]{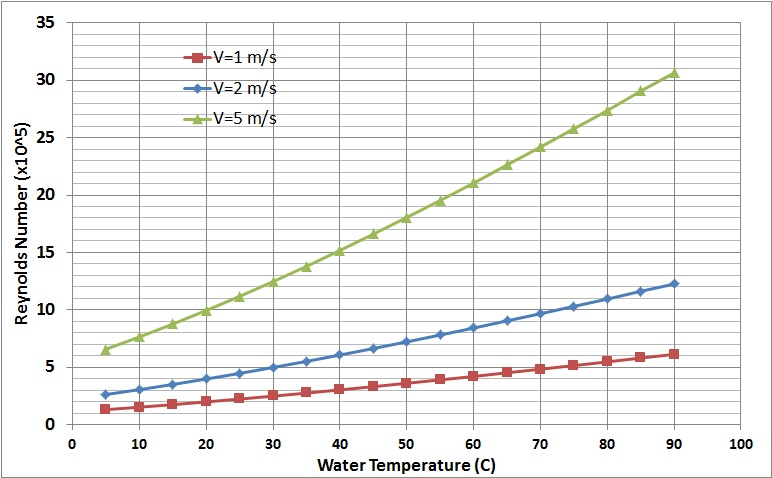}
\caption{${\rm Re}_D$ for water in function of $T$, for $D=\unit[5]{cm}$ and $V=1, 2, 5\,{m/s}$ as given by Eq.(\ref{Reynolds}).}
\label{fig:ReynTh2o}
\end{figure}

Consider a cylindrical pipe of external diameter $D$. Some fluid is flowing across the pipe, with velocity $V$ in proximity of its surface. The Reynolds Number is given by the following expression,
\begin{equation}
\label{Reyndef}
{\rm Re}_D=\dfrac{\rho VD}{\mu}=\dfrac{VD}{\nu}.
\end{equation}
Consider first dry air. Since the kinematic viscosity $\nu(T)$ (or, equivalently, the ratio $\mu/\rho$) is a function of the temperature, namely \cite{Airprop}
\be\label{Airnu}
\nu(T)=\left( \frac{2.409\times 10^8}{T^{3/2}}+\frac{2.6737\times 10^{10}}{T^{5/2}} \right)^{-1}\,\left[\frac{m^2}{s}\right]\,,
\ee
where $[T]=[K]$, the Reynolds number can be written explicitly as ${\rm Re}_D(T)$:
\bea\label{Reynolds}
{\rm Re}_D(T)=\frac{VD}{T^{3/2}}\left( 2.409+\frac{267.37}{T} \right)\times 10^8\,.
\eea
The dependence on $V$ and on the diameter $D$ is trivial, however how the Reynolds number changes in function of the temperature is not. This is plotted in Fig.\ref{fig:ReynT}.
Table \ref{table:ReynTWater} and Table \ref{table:ReynTAir} list several values of ${\rm Re}_D(T)$. Table \ref{table:ReynVWater} and Table \ref{table:ReynVAir} give values of $\rm Re (V)$.

We need to find the convection coefficient which depends on the Reynolds number in order to calculate the heat transfer rate from the plate to the flowing fluid. For that purpose we need the Nusselt Number which contains the convection coefficient and is calculated via the Reynolds number. However the flow over a cylinder (or a sphere) has to become turbulent after the impact. Thus the Nusselt number evolves in function of the position of fluid particles beside the impact point.

Therefore, to calculate the \emph{average} heat transfer coefficient we use the Churchill-Bernstein correlation for the average Nusselt number \cite{CB,IncdeVitt},
\be
{\rm Nu}_{cyl}=\frac{hD}{k}=0.3+\frac{0.62{\rm Re}^{1/2}{\rm Pr}^{1/3}}{[1+(0.4/{\rm Pr})^{2/3}]^{1/4}}
\left[1+\left(\dfrac{{\rm Re}}{28200}\right)^{5/8}\right]^{4/5}.
\ee
The convection coefficient then can be rewritten as a function of the variables of interest as follows,
\bea\label{hvDnu}
h(V,D,\nu)&=&k(T)\left\{ 
\frac{0.3}{D}+0.62\frac{{\rm Pr(T)}^{1/3}}{\left[1+\left(0.4/{\rm Pr(T)}\right)^{2/3})\right]^{1/4}}
\right.
\nonumber\\
&\times&
\left.
\sqrt{\frac{V}{\nu(T)D}}
\left[1+\left(\frac{V D}{28200\nu(T)}\right)^{5/8}\right]^{4/5}
\right\}\,,
\eea
where $k$ is the thermal conductivity of the fluid, and the dependence on $T$ is explicit. The above formula is common to both air and water.

	The Prandtl number is obtained from ${\rm Pr}=\nu/\alpha$, where the thermal diffusivity is written as \cite{Airprop}
	\be
	\label{Airalpha}
	\alpha(T)=-4.3274+4.1190\times 10^{-2}T+1.5556\times 10^{-4}T^2\,\left[10^{-6}\frac{m^2}{s} \right]\,,
	\ee
and the thermal conductivity is retrieved via Sutherland's equation (Reid, 1966) \cite{Airprop}:
\be
\label{Airk}
k(T)=\frac{2.3340T^{3/2}}{164.54+T}\qquad\left[10^{-3} \frac{W}{mK} \right]\,.
\ee
By substituting the above correlations into Eq.(\ref{hvDnu}), one therefore obtains the explicit dependence of $h$ on the film temperature $T=\left(T_{s}+T_{\infty}\right)/2$ alone, with $V$ and $D$ fixed. The diameter of the pipe and the fluid velocity are indeed the parameters easily controlled.

\begin{figure}[ht]
\begin{center}
\includegraphics[width=0.9\textwidth]{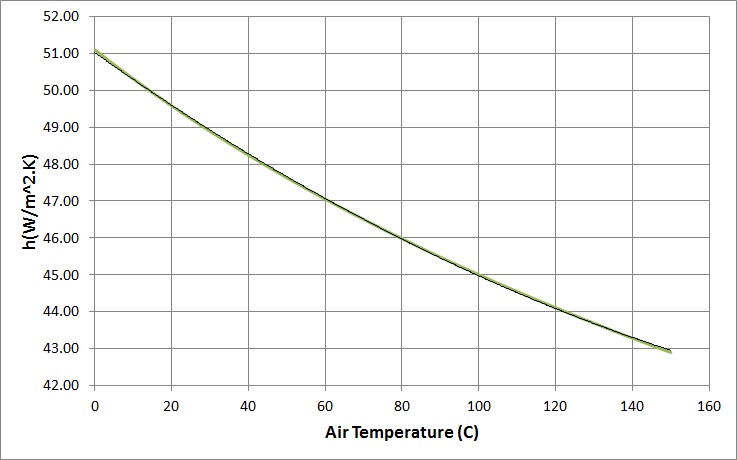}
\caption{Air convection coefficient in function of temperature for $D=\unit[5]{cm}$ and $V=\unit[5]{m/s}$.}
\label{fig:hTAir}
%
\vspace{3mm}

\includegraphics[width=0.9\textwidth]{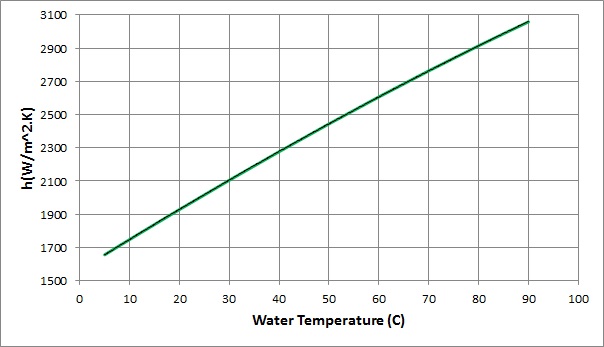}
\caption{Water convection coefficient in function of temperature for $D=\unit[5]{cm}$ and $V=\unit[0.2]{m/s}$.}
\label{fig:hTWater}
\end{center}
\end{figure}

Since the explicit form of $h(T)$ is very involved, due to Eqs.(\ref{Airnu}), (\ref{Airalpha}) and (\ref{Airk}), we prefer to plot the values for $0^oC<T<150^oC$ and find the interpolation curve
\be\label{hair}
h_{air}(T)=10^{-4}T^2-0.0739T+51.022\qquad\left[\frac{W}{m^2K} \right]\,,
\ee
where $[T]=[K]$. This is shown in Fig.\ref{fig:hTAir}, that gives a direct relation between film temperature and convection coefficient.
Note that the heat transfer coefficient in (\ref{hair}) has a minimum for $T\sim \unit[370]{^\circ C}$, then starts raising again. 

Therefore we can conclude that for air flowing through a single cylinder, heat transfer becomes less efficient with increasing air temperature, for any acceptable value of $T$.


	Following exactly the same reasoning, we now compute the heat transfer coefficient for water. This is slightly more difficult than for air, due to the lack of interpolation curves for water available in the literature. Therefore we compute them as follows.
		
	The expression $\nu(T)$ for the kinematic viscosity can be found from the following formulas for the dynamic viscosity \cite{waterprop1}
	\be\label{muh2o}
	\mu(T)=(280.68\,T_*^{-1.9}+511.45\,T_*^{-7.7}+61.131\,T_*^{-19.6}+0.45903\,T_*^{-40}
	)
	\left[10^{-6}\frac{kg}{ms}\right],
	\ee
where $T_*=T/300K$, and for the density (with $[T]=[C]$)
\be\label{rhoh2o}
\rho(T)=2\times 10^{-5}\,T^3-0.0063\,T^2+0.0266\,T+999.98
\qquad \left[\frac{kg}{m^3}\right]\,.
\ee
To find the analytical expression for $\nu(T)$, we then compute the values by using (\ref{muh2o}) and (\ref{rhoh2o}). These are consistent with e.g. \cite{IncdeVitt,waterprop1}. Interpolating the resulting curve gives
\be\label{nuh2o}
\nu(T)=3\times10^{-8}\,T^4 - 9\times10^{-6}\,T^3 + 0.001\,T^2 - 0.0552\,T + 1.7796
\qquad
\left[10^{-6}\frac{m^2}{s}\right]\,,
\ee
where the temperature is in degree Celsius. The corresponding Reynolds number values are plotted in Fig.\ref{fig:ReynTh2o}.

	The thermal diffusivity $\alpha$ for water is evaluated by means of $\alpha=k/(\rho c_p)$, and results in the following correlation formula,
	\be
	\alpha(T)=-2\times10^{-5}\,T^2 + 0.0049\,T + 1.3491
	\qquad \left[10^{-7} \frac{m^2}{s}\right].
	\ee
	This is obtained from the thermal conductivity \cite{waterprop2}
	\be\label{kh2o}
	k(T)=0.5678+1.8774\times 10^{-3}\,T-8.179\times 10^{-6}\,T^2+5.6629\times 10^{-9}\,T^3\,\left[\frac{W}{mK}\right],
	\ee
	and from the following expression for the specific heat \cite{IncdeVitt,physicalvalues}
	\bea
	c_p(T)&=& -3\times10^{-8}\,T^5 + 10^{-5}\,T^4 - 0.0014\,T^3
	\nonumber\\
	&& + 0.0978\,T^2 - 3.2467\,T + 4217.7 \qquad \left[ \frac{J}{kgK} \right].
	\eea
	Similarly as before, we obtain the Prandtl number for water from ${\rm Pr}=\nu/\alpha$, whose values are cross-checked also with \cite{IncdeVitt}. Then we insert (\ref{nuh2o}) and (\ref{kh2o}) in Eq.(\ref{hvDnu}).
	
	For water flowing at $V=0.2\,m/s$ over a cylindrical pipe with $D=5\,cm$, we find for the temperature range $5^oC<T<90^oC$
	\be
	h(T)=-0.0252\,T^2 + 18.925\,T + 1560.9\qquad \left[\frac{W}{m^2K}\right],
	\ee
	where $[T]=[C]$. The above heat transfer coefficient is plotted in Fig.\ref{fig:hTWater}. Note that also here there is a maximum, but it occurs for $T\sim\unit[375]{^\circ C}$, thus very well above the boiling point. Thus heat transfer between a water flow and a cylindrical bar increases with the water temperature for any realistic $T$ values.

\begin{figure}[hp]
\centering
\includegraphics[width=0.8\textwidth]{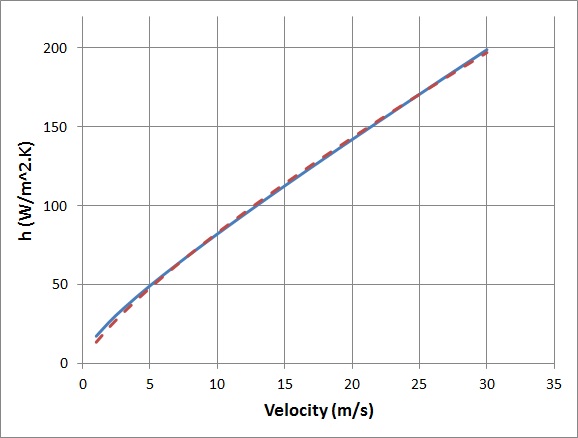}
\caption{$h$ for air in function of $V$, for $D=\unit[5]{cm}$ and $T_f=\unit[15]{^\circ C}$, as given by Eq.(\ref{hvDnu}). The fit curve (\ref{hinterpairv}) is dashed.
}
\label{fig:hinterpairv}
\vspace{3mm}


\centering
\includegraphics[width=0.8\textwidth]{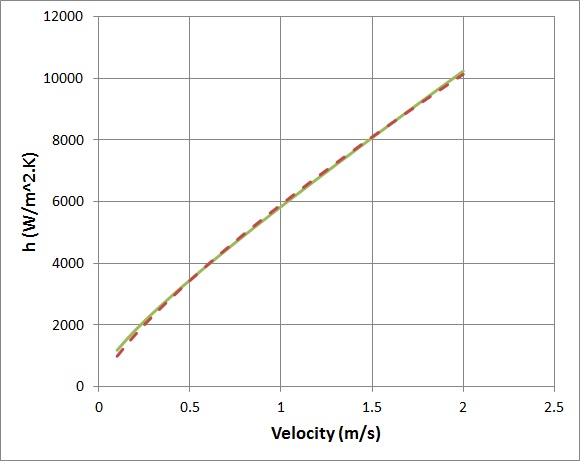}
\caption{$h$ for water in function of $V$, for $D=\unit[5]{cm}$ and $T_f=\unit[15]{^\circ C}$, as given by Eq.(\ref{hvDnu}). The fit curve (\ref{hinterph2ov}) is dashed.
}
\label{fig:hinterph2ov}
\end{figure}

	On the other hand, graphs of Eq.(\ref{hvDnu}) for air and water, where $h$ is a function of the fluid velocity with fixed $T=\unit[15]{^\circ C}$ and $D=5\,cm$, can be found in Figs \ref{fig:hinterpairv} and \ref{fig:hinterph2ov}. The according interpolation curves,
\be\label{hinterpairv}
h_{water}(V)=5901.71\times V^{0.779}\,,
\ee
\be\label{hinterph2ov}
h_{air}(V)=13.4712\times V^{0.789}\,,
\ee
are obtained with the least square method, as described in the Appendix.


	For both fluids, the dependence of Re and $h$ on the fluid velocity $V$ is trivial. However, the temperature influences the Reynolds number and heat transfer coefficient in opposite ways. For a liquid like water, the viscosity decreases at higher temperatures, thus its ability to transfer heat by convection is improved. On the other hand, since air is a gas, it has increasing viscosity with temperature, see Eq.(\ref{Airnu}). This hinders the heat transfer as seen in Eq.(\ref{hair}), as the dominant term is $\propto -T$.

Fig \ref{fig:hTAir} and Fig \ref{fig:hTWater} give the evolution of $h$ in function of the temperature when $V$ and $D$ are constant. The according values are given in Tables \ref{table:hvinfAir} and \ref{table:hvinfWater}.

\begin{figure}[ht]
\begin{center}
\includegraphics[width=0.9\textwidth]{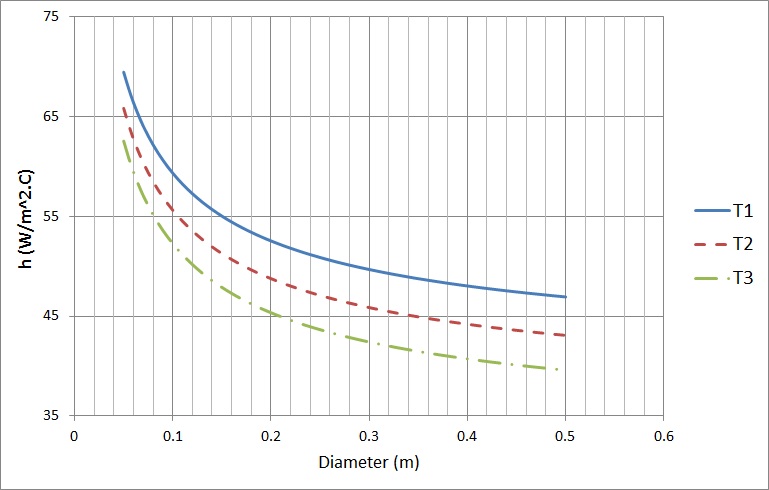}
\caption{Air convection coefficient in function of the diameter for $V=\unit[8]{m/s}$. The film temperatures are $T_1=\unit[15]{^\circ C}$, $T_2=\unit[50]{^\circ C}$ and $T_3=\unit[95]{^\circ C}$. }
\label{fig:hDAir}
%

\vspace{3mm}

\includegraphics[width=0.9\textwidth]{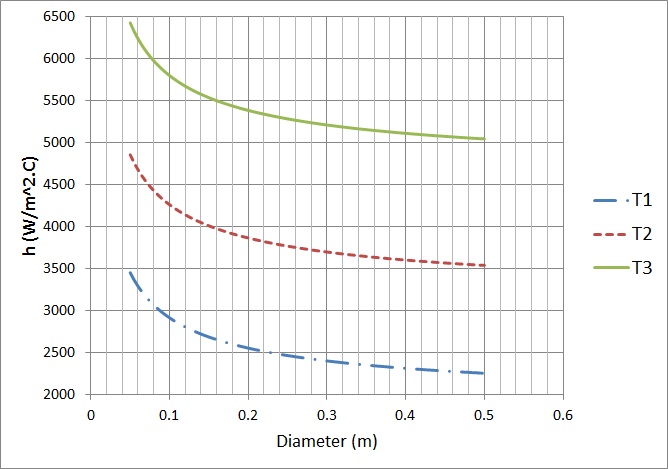}
\caption{Water convection coefficient in function of the diameter for $V=\unit[0.5]{m/s}$. The film temperatures are $T_1=\unit[15]{^\circ C}$, $T_2=\unit[50]{^\circ C}$ and $T_3=\unit[95]{^\circ C}$. }
\label{fig:hDWater}
\end{center}
\end{figure}

	The diameter also has a relevant influence on $h$. Fig \ref{fig:hDAir} and Fig \ref{fig:hDWater} show that the convection coefficient is inversely proportional to the diameter of the cylinder, and quite strongly so (power law). Indeed for a larger diameter the flow is more likely to become turbulent earlier, closer to the impact point. The ordered fluid motion in a laminar flow transfers heat more efficiently. Also $\rm Re$ (which indicates the nature of the flow) is a function of $D$.
	
	As explained above, the temperature still has an influence.
	However, it is less significant than the diameter, since $h$ is mostly linearly dependent on $T$. This is especially evident for air, whose properties do not change appreciably in this particular case ($\unit[0]{^\circ C}$ to $\unit[100]{^\circ C}$).

\begin{figure}[ht]
\centering
\includegraphics[width=0.8\textwidth]{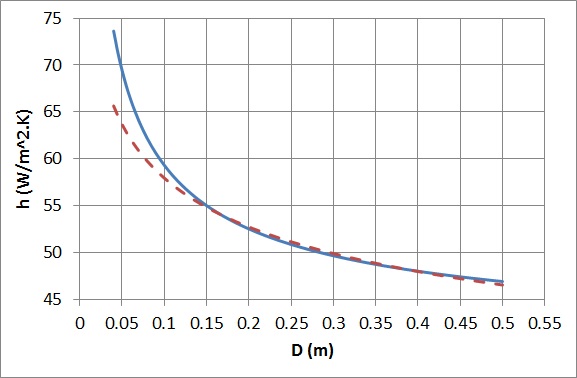}
\caption{Air convection coefficient in function of the diameter for $L=\unit[1]{m}$, $V=\unit[8]{m/s}$ and $T=\unit[15]{^\circ C}$.
The fit curve (\ref{fitaD}) is dashed.
}
\label{fig:hDAir2}
\end{figure}

\begin{figure}[ht]
\begin{center}
\includegraphics[width=0.8\textwidth]{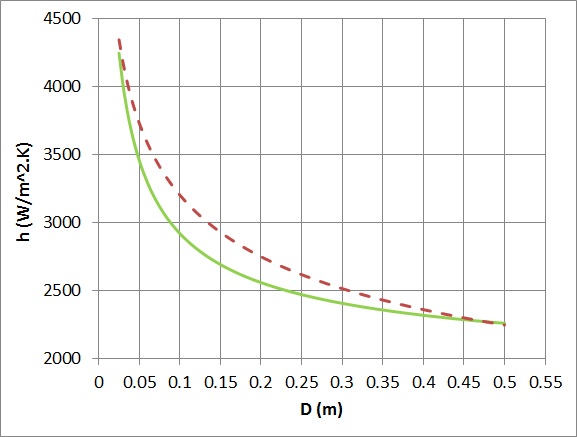}
\caption{Water convection coefficient in function of the diameter for $L=\unit[1]{m}$, $V=\unit[8]{m/s}$ and $T=\unit[15]{^\circ C}$.
The fit curve (\ref{fitwD}) is dashed.
}
\label{fig:hDWater2}
\end{center}
\end{figure}

We can also calculate the expressions of these functions at $\unit[15]{^\circ C}$, and obtain
\bea
h_{air}(D)&=&42.37\times D^{-0.136}\,,\label{fitaD}
\\
h_{water}(D)&=&1838.44\times D^{-0.223}\,.\label{fitwD}
\eea
These are plotted in Fig \ref{fig:hDAir2} and Fig \ref{fig:hDWater2}, where the convection coefficient decreases with a low derivative, almost logarithmically. This makes sense because the Nusselt number is expressed as a power function of the Reynolds number which is proportional to the diameter \emph{and} the velocity. Therefore the convection coefficient is very sensitive to a diameter increase, but only when $D$ is rather small. Above $\unit[10]{cm}$ to $\unit[20]{cm}$, the diameter has not much influence; the flow is mainly laminar over the cylinder and the heat transfer coefficient is small.

\clearpage

\section*{Tables and Plots}

\begin{table}[ht]
\begin{minipage}[b]{0.50\linewidth}\centering
\begin{tabular}{c|c|c|c|}
\cline{2-4}
 & $V_{1}$ & $V_{2}$ & $V_{3}$ \\ \cline{2-4}
 & \unit[5]{m/s}	& \unit[10]{m/s} & \unit[15]{m/s} \\ \hline
\multicolumn{1}{|c|}{$T(^\circ$C)} & \multicolumn{3}{|c|}{Reynolds Number ( $\times 10^5$)} \\ \hline
\multicolumn{1}{|c|}{0.01} & 1.39	& 2.79	& 4.18
 \\ \hline
\multicolumn{1}{|c|}{5} & 1.88 & 3.76 & 4.94 \\ \hline
\multicolumn{1}{|c|}{10} & 1.91 & 3.82 & 5.74 \\ \hline
\multicolumn{1}{|c|}{15} & 2.19 & 4.39 & 6.58 \\ \hline
\multicolumn{1}{|c|}{20} & 2.49 & 4.98 & 7.47 \\ \hline
\multicolumn{1}{|c|}{25} & 2.80 & 5.59 & 8.39 \\ \hline
\multicolumn{1}{|c|}{30} & 3.12 & 6.24 & 9.36 \\ \hline
\multicolumn{1}{|c|}{35} & 3.45 & 6.90 & 10.35 \\ \hline
\multicolumn{1}{|c|}{40} & 3.80 & 7.60 & 11.39 \\ \hline
\multicolumn{1}{|c|}{45} & 4.15 & 8.31 & 12.46 \\ \hline
\multicolumn{1}{|c|}{50} & 4.52 & 9.03 & 13.55 \\ \hline
\multicolumn{1}{|c|}{55} & 4.89 & 9.77 & 14.66 \\ \hline
\multicolumn{1}{|c|}{60} & 5.26 & 10.53 & 15.79 \\ \hline
\multicolumn{1}{|c|}{65} & 5.66 & 11.32 & 16.98 \\ \hline
\multicolumn{1}{|c|}{70} & 6.05 & 12.10 & 18.15 \\ \hline
\multicolumn{1}{|c|}{75} & 6.45 & 12.89 & 19.34 \\ \hline
\multicolumn{1}{|c|}{80} & 6.84 & 13.69 & 20.53 \\ \hline
\multicolumn{1}{|c|}{85} & 7.27 & 14.54 & 21.80 \\ \hline
\multicolumn{1}{|c|}{90} & 7.66 & 15.32 & 22.98 \\ \hline
\multicolumn{1}{|c|}{95} & 8.09 & 16.19 & 24.28 \\ \hline
\end{tabular}
\caption{Reynolds numbers for water in function of $T$ and $V$ for $D=\unit[5]{cm}$.}
\label{table:ReynTWater}
\end{minipage}
\hspace{0.5cm}
\begin{minipage}[b]{0.50\linewidth}
\centering
\begin{tabular}{c|c|c|c|}
\cline{2-4}
 & $V_{1}$ & $V_{2}$ & $V_{3}$ \\ \cline{2-4}
 & \unit[5]{m/s}	& \unit[10]{m/s} & \unit[15]{m/s} \\ \hline
\multicolumn{1}{|c|}{$T$($^\circ$C)} & \multicolumn{3}{|c|}{Reynolds Number ( $\times 10^4$)} \\ \hline
\multicolumn{1}{|c|}{0} & 1.88 &	3.75 &	5.63\\ \hline
\multicolumn{1}{|c|}{5} & 1.82 &	3.63 &	5.45
 \\ \hline
\multicolumn{1}{|c|}{10} & 1.76	& 3.52	& 5.28
 \\ \hline
\multicolumn{1}{|c|}{15} & 1.71	& 3.41	& 5.12
 \\ \hline
\multicolumn{1}{|c|}{20} & 1.65	& 3.31	& 4.96
 \\ \hline
\multicolumn{1}{|c|}{25} & 1.61	& 3.21	& 4.82
 \\ \hline
\multicolumn{1}{|c|}{30} & 1.56	& 3.12	& 4.68
 \\ \hline
\multicolumn{1}{|c|}{35} & 1.51	& 3.03	& 4.54
 \\ \hline
\multicolumn{1}{|c|}{40} & 1.47	& 2.94	& 4.42
 \\ \hline
\multicolumn{1}{|c|}{45} & 1.43	& 2.86	& 4.29
 \\ \hline
\multicolumn{1}{|c|}{50} & 1.39	& 2.79	& 4.18
 \\ \hline
\multicolumn{1}{|c|}{55} & 1.36	& 2.71	& 4.07
 \\ \hline
\multicolumn{1}{|c|}{60} & 1.32	& 2.64	& 3.96
 \\ \hline
\multicolumn{1}{|c|}{65} & 1.29	& 2.57	& 3.86
 \\ \hline
\multicolumn{1}{|c|}{70} & 1.25	& 2.51	& 3.76
 \\ \hline
\multicolumn{1}{|c|}{75} & 1.22	& 2.45	& 3.67
 \\ \hline
\multicolumn{1}{|c|}{80} & 1.19	& 2.39	& 3.58
 \\ \hline
\multicolumn{1}{|c|}{85} & 1.16	& 2.33	& 3.49
 \\ \hline
\multicolumn{1}{|c|}{90} & 1.14	& 2.27	& 3.41
 \\ \hline
\multicolumn{1}{|c|}{95} & 1.11	& 2.22	& 3.33
 \\ \hline
\multicolumn{1}{|c|}{100} & 1.08 &	2.17 &	3.25
 \\ \hline
\multicolumn{1}{|c|}{105} & 1.06 &	2.12 & 3.18

 \\ \hline
%

\end{tabular}
\caption{Reynolds numbers for air in function of $T$ and $V$ for $D=\unit[5]{cm}$.}
\label{table:ReynTAir}
\end{minipage}
\end{table}

\begin{table}[ht]
\centering
\begin{minipage}[b]{0.45\linewidth}\centering
\begin{tabular}{c|c|c|c|}
\cline{2-4}
 & $T_1$ & $T_2$ & $T_3$ \\ \cline{2-4}
 & \unit[5]{$^\circ$ C} & \unit[50]{$^\circ$ C} & \unit[95]{$^\circ$ C} \\ \hline
\multicolumn{1}{|c|}{$V$} & \multicolumn{3}{|c|}{Reynolds Number ($\times 10^4$)} \\ \hline
\multicolumn{1}{|c|}{1} & 3.29 & 9.03 & 16.19 \\ \hline
\multicolumn{1}{|c|}{2} & 6.58 & 18.06 & 32.37 \\ \hline
\multicolumn{1}{|c|}{3} & 9.87 & 27.10 & 48.56 \\ \hline
\multicolumn{1}{|c|}{4} & 13.16 & 36.13 & 64.75 \\ \hline
\multicolumn{1}{|c|}{5} & 16.45 & 45.16 & 80.93 \\ \hline
\multicolumn{1}{|c|}{6} & 19.74 & 54.19 & 97.12 \\ \hline
\multicolumn{1}{|c|}{7} & 23.03 & 63.22 & 113.31 \\ \hline
\multicolumn{1}{|c|}{8} & 26.33 & 72.26 & 129.49 \\ \hline
\multicolumn{1}{|c|}{9} & 29.62 & 81.29 & 145.68 \\ \hline
\multicolumn{1}{|c|}{10} & 32.91 & 90.32 & 161.87 \\ \hline
\multicolumn{1}{|c|}{11} & 36.20 & 99.35 & 178.06 \\ \hline
\multicolumn{1}{|c|}{12} & 39.49 & 108.38 & 194.24 \\ \hline
\multicolumn{1}{|c|}{13} & 42.78 & 117.42 & 210.43 \\ \hline
\multicolumn{1}{|c|}{14} & 46.07 & 126.45 & 226.62 \\ \hline
\multicolumn{1}{|c|}{15} & 49.36 & 135.48 & 242.80 \\ \hline
\multicolumn{1}{|c|}{16} & 52.66 & 144.51 & 258.99 \\ \hline
\multicolumn{1}{|c|}{17} & 55.95 & 153.54 & 275.18 \\ \hline
\multicolumn{1}{|c|}{18} & 59.24 & 162.58 & 291.36 \\ \hline
\multicolumn{1}{|c|}{19} & 62.53 & 171.61 & 307.55 \\ \hline
\multicolumn{1}{|c|}{20} & 65.82 & 180.64 & 323.74 \\ \hline
\end{tabular}
\caption{{\rm Re}($V,T$) for water, $D=\unit[5]{cm}$ and $[V]$=[\unit{m/s}].}
\label{table:ReynVWater}
\end{minipage}
\hspace{0.5cm}
\begin{minipage}[b]{0.45\linewidth}
\centering
\begin{tabular}{c|c|c|c|}
\cline{2-4}
 & $T_1$ & $T_2$ & $T_3$ \\ \cline{2-4}
 & \unit[5]{$^\circ$ C} & \unit[50]{$^\circ$ C} & \unit[95]{$^\circ$ C} \\ \hline
\multicolumn{1}{|c|}{$V$} & \multicolumn{3}{|c|}{Reynolds Number ($\times 10^3$)} \\ \hline
\multicolumn{1}{|c|}{1} & 3.62 & 2.78 & 2.27 \\ \hline
\multicolumn{1}{|c|}{2} & 7.24 & 5.56 & 4.54 \\ \hline
\multicolumn{1}{|c|}{3} & 10.85 & 8.34 & 6.82 \\ \hline
\multicolumn{1}{|c|}{4} & 14.47 & 11.13 & 9.09 \\ \hline
\multicolumn{1}{|c|}{5} & 18.09 & 13.91 & 11.36 \\ \hline
\multicolumn{1}{|c|}{6} & 21.71 & 16.69 & 13.63 \\ \hline
\multicolumn{1}{|c|}{7} & 25.32 & 19.47 & 15.90 \\ \hline
\multicolumn{1}{|c|}{8} & 28.94 & 22.25 & 18.17 \\ \hline
\multicolumn{1}{|c|}{9} & 32.56 & 25.03 & 20.45 \\ \hline
\multicolumn{1}{|c|}{10} & 36.17 & 27.82 & 22.72 \\ \hline
\multicolumn{1}{|c|}{11} & 39.79 & 30.60 & 24.99 \\ \hline
\multicolumn{1}{|c|}{12} & 43.41 & 33.38 & 27.26 \\ \hline
\multicolumn{1}{|c|}{13} & 47.03 & 36.16 & 29.53 \\ \hline
\multicolumn{1}{|c|}{14} & 50.64 & 38.94 & 31.80 \\ \hline
\multicolumn{1}{|c|}{15} & 54.26 & 41.72 & 34.07 \\ \hline
\multicolumn{1}{|c|}{16} & 57.88 & 44.50 & 36.34 \\ \hline
\multicolumn{1}{|c|}{17} & 61.50 & 47.29 & 38.62 \\ \hline
\multicolumn{1}{|c|}{18} & 65.11 & 50.07 & 40.89 \\ \hline
\multicolumn{1}{|c|}{19} & 68.73 & 52.85 & 43.16 \\ \hline
\multicolumn{1}{|c|}{20} & 72.35 & 55.63 & 45.43 \\ \hline
\end{tabular}
\caption{{\rm Re}($V,T$) for air, $D=\unit[5]{cm}$ and $[V]$=[\unit{m/s}].}
\label{table:ReynVAir}
\end{minipage}
\end{table}

\begin{table}[ht]
\begin{minipage}[b]{0.45\linewidth}\centering
\begin{tabular}{|c|c|}
\hline
$T$($^\circ$C )& $h$(W/m$^2$C)  \\ \hline
5 & 50.70 \\ \hline
20 & 49.58 \\ \hline
30 & 48.89 \\ \hline
40 & 48.24 \\ \hline
50 & 47.63 \\ \hline
60 & 47.05 \\ \hline
70 & 46.50 \\ \hline
80 & 45.98 \\ \hline
90 & 45.48 \\ \hline
100 & 45.00 \\ \hline
\end{tabular}
\caption{$h$ for air, with $V=5\,m/s$ and $D=\unit[5]{cm}$.}
\label{table:hTAir}
\end{minipage}
\hspace{0.5cm}
\begin{minipage}[b]{0.45\linewidth}
\centering
\begin{tabular}{|c|c|}
\hline
$T$($^\circ$C ) & $h$(W/m$^2$C)  \\ \hline
5 & 1656.02 \\ \hline
10 & 1747.85 \\ \hline
20 & 1928.34 \\ \hline
30 & 2104.71 \\ \hline
40 & 2276.80 \\ \hline
50 & 2444.29 \\ \hline
60 & 2606.70 \\ \hline
70 & 2763.52 \\ \hline
80 & 2914.27 \\ \hline
90 & 3058.54 \\ \hline
\end{tabular}
\caption{$h$ for water, with $V=0.2\,m/s$ and $D=\unit[5]{cm}$.}
\label{table:hTWater}
\end{minipage}
\end{table}

\begin{table}[ht]
\begin{minipage}[b]{0.45\linewidth}\centering
\begin{tabular}{|c|c|}
\hline
$V$(m/s) & $h$(W/m$^2$C)  \\ \hline
2 & 26.59 \\ \hline
4 & 42.20 \\ \hline
6 & 56.21 \\ \hline
8 & 69.43 \\ \hline
10 & 82.16 \\ \hline
12 & 94.55 \\ \hline
14 & 106.69 \\ \hline
16 & 118.63 \\ \hline
18 & 130.41 \\ \hline
20 & 142.06 \\ \hline
\end{tabular}
\caption{Convection Coefficient for air at $T_f=15^{\protect\circ}$C in function of $V$ for $D=\unit[5]{cm}$.}
\label{table:hvinfAir}
\end{minipage}
\hspace{0.5cm}
\begin{minipage}[b]{0.45\linewidth}
\centering
\begin{tabular}{|c|c|}
\hline
$V$(m/s) & $h$(W/m$^2$C)  \\ \hline
0.2 & 1827.15 \\ \hline
0.4 & 2940.46 \\ \hline
0.6 & 3950.35 \\ \hline
0.8 & 4909.08 \\ \hline
1 & 5836.18 \\ \hline
1.2 & 6741.40 \\ \hline
1.4 & 7630.38 \\ \hline
1.6 & 8506.76 \\ \hline
1.8 & 9373.0 \\ \hline
2 & 10230.87 \\ \hline
\end{tabular}
\caption{Convection Coefficient for water at $T_f=15^{\protect\circ}$C in function of $V$ for $D=\unit[5]{cm}$.}
\label{table:hvinfWater}
\end{minipage}
\end{table}

\clearpage


%


\begin{figure}[ht]
\begin{center}
\includegraphics[width=0.8\textwidth]{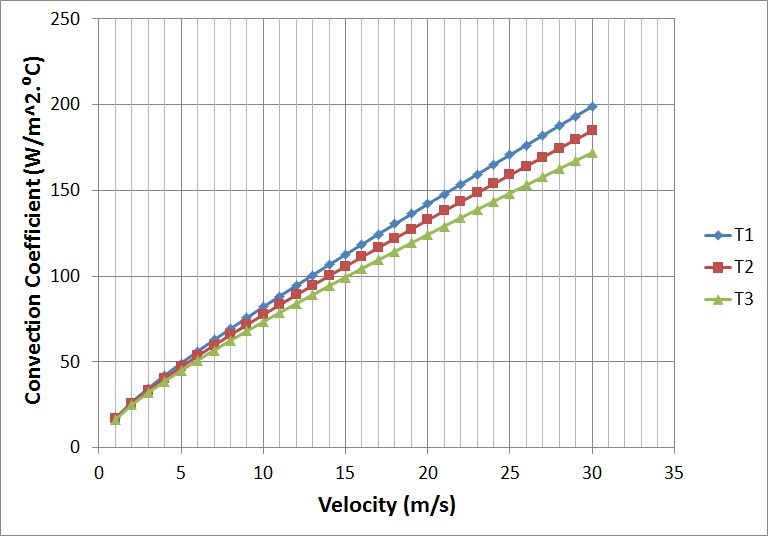}
\caption{Air convection coefficient in function of the velocity for $T_1=\unit[15]{^\circ C}$, $T_2=\unit[50]{^\circ C}$ and $T_3=\unit[95]{^\circ C}$. $D=5\,cm$.}
\label{fig:hVinfAir}
%

\vspace{3mm}

\includegraphics[width=0.8\textwidth]{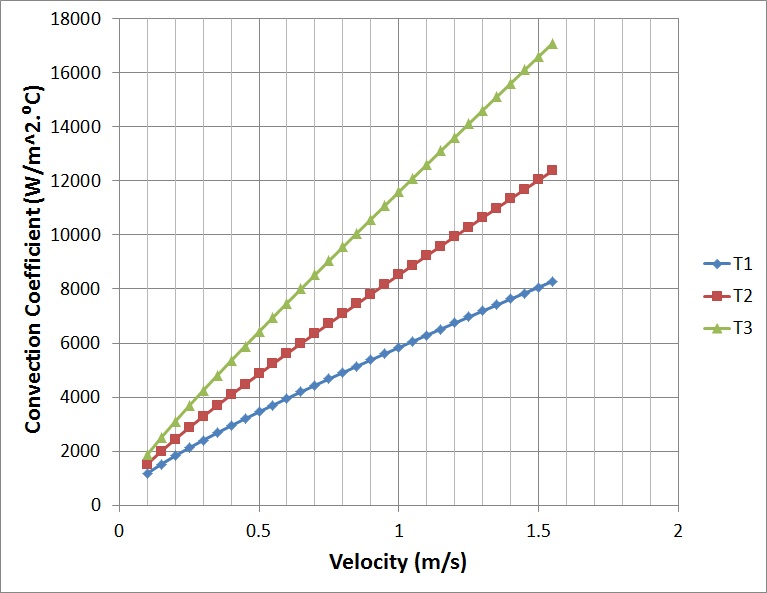}
\caption{Water convection coefficient in function of the velocity for $D=\unit[5]{cm}$.
$T_1=\unit[15]{^\circ C}$, $T_2=\unit[50]{^\circ C}$ and $T_3=\unit[95]{^\circ C}$.}
\label{fig:hVinfWater}
\end{center}
\end{figure}

\clearpage

\section{Convection inside a pipe}


Let us consider a cylindrical pipe with constant surface temperature or constant surface heat flux. A fluid flows inside the pipe,
with properties evaluated at the bulk temperature\footnote{A common method, which we follow here, is to define $T=(T_i+T_e)/2$, where $T_i$ is the inlet temperature and $T_e$ the exit (or supply) temperature.}. In this study we choose dry air at 1 atm (flow inside a ventilation duct), as well as R-717 (namely ammonia), a refrigerant commonly used e.g. in the cooling pipes under ice hockey rinks. This particular situation is especially pertinent to the case studied in this section.

\begin{figure}[t]
\centering
\includegraphics[width=0.9\textwidth]{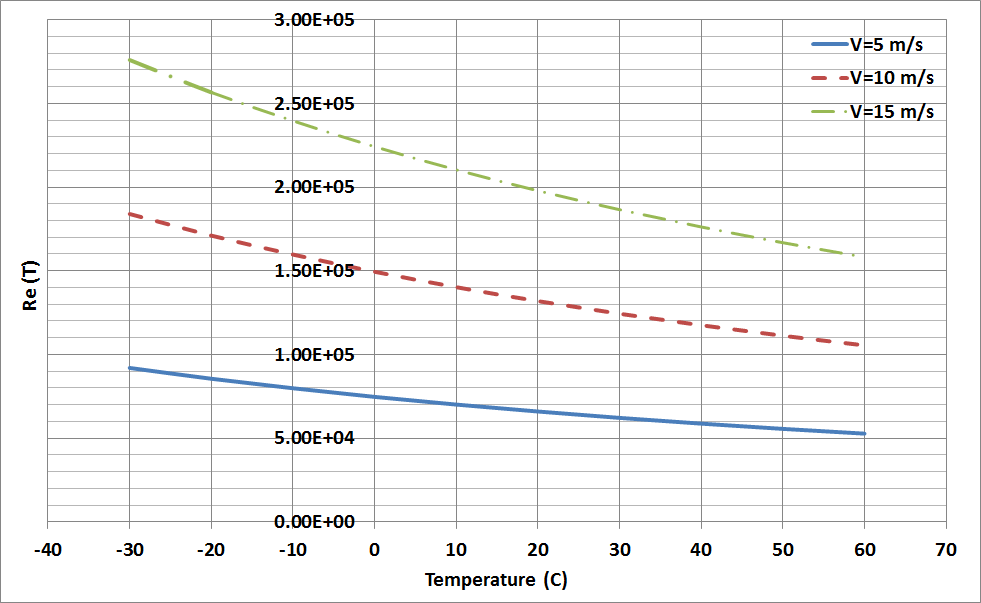}
\caption{Reynolds number for air in function of the bulk temperature, for $V_\infty=5,10,15\,m/s$ and $D=20\,cm$.}
\label{fig:RevsT}
\vspace{3mm}
\includegraphics[width=0.9\textwidth]{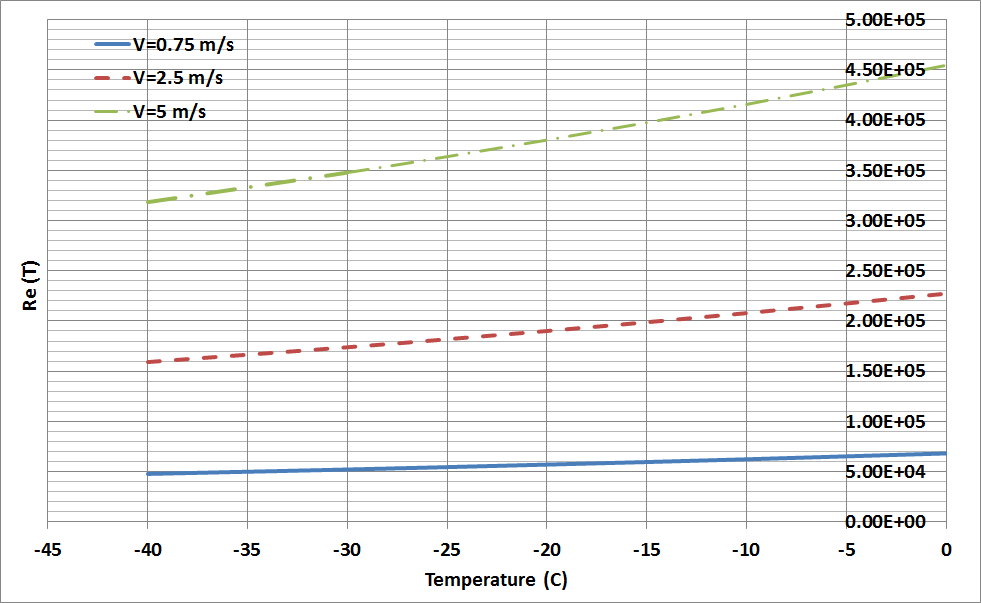}
\caption{Reynolds number for ammonia in function of the bulk temperature, for $V_\infty=0.75,2.5,5\,m/s$ and $D=27\,mm$.}
\label{fig:RevsTamm}

\end{figure}

Recall the definition (\ref{Reyndef}) for the Reynolds number, which still holds here.
The characteristic length is in this case the internal diameter $D$
\be
{\rm Re=}\frac{V_{\infty}D}{\nu}.
\ee 
Air properties are discussed in the previous section, Eqs.(\ref{Airnu}), (\ref{Airalpha}) and (\ref{Airk}). Values of the Reynolds number for the temperature range $0^\circ C<T<105^\circ C$ are listed in Table \ref{table:ReynTAir} for a cylinder of external diameter $D=5\,cm$.
	Since here we discuss a $D=20\,cm$ duct, it is enough to multiply those values by a factor of 4. Values of ${\rm Re}(T,V_\infty)$ for a different temperature range are given instead in Table \ref{table:ReynTAirInside} and in Table \ref{table:ReynVAirInside}. The according plot is found in Fig.\ref{fig:RevsT}.

	Consider now saturated ammonia. The thermophysical properties are as follows \cite{Ammoniaprop},
\bea
k(T) &=& -0.0023\,T + 0.539\qquad\left[ \frac{W}{mK}\right],\\
\rho(T)&=& -3\times 10^{-5}\,T^3 - 0.0011\,T^2 - 1.362\,T + 638.2 \; \left[ \frac{kg}{m^3}\right],\\
\nu(T)&=& 10^{-4}\, T^2 - 0.0262\,T + 2.9748 \qquad \left[ 10^{-7}\frac{m^2}{s}\right],
\eea
the Prandtl number is
\be
{\rm Pr}(T)= 5\times10^{9}\,T^4 - 3\times10^{-7}\,T^3 + 5\times10^{-5}\,T^2 - 0.0087\,T + 1.6225\,,
\ee
and the specific heat capacity holds as
\be
c_p(T)= 3\times 10^{-5}\,T^4 - 0.0015\,T^3 + 0.0522\,T^2 + 6.7534\,T + 4609.8\; 
\left[ \frac{J}{kgK}\right].
\ee
	In all the equations above, $[T]=[C]$. These give the tables \ref{table:ReynTRef} and \ref{table:ReynVRef}, and the plot in Fig.\ref{fig:RevsTamm}.
	  From Figs.\ref{fig:RevsT} and \ref{fig:RevsTamm}, one can infer that the Reynolds number is linearly dependent on temperature for both air and ammonia (with negative derivative in the case of air).
	Let us stress that this is an example of a result which does not usually appear in textbooks.

\subsection{Heat transfer coefficient}
	The convection coefficient is obtained via the Nusselt number $\rm Nu$, through
\be
h=\frac{k}{D}{\rm Nu}\,.
\label{eq:hNu}
\ee
$\rm Nu$ depends on the nature of the flow. This is determined by means of the Reynolds number as follows:
\be
\begin{cases}
{\rm Re}<2\,300 & \qquad laminar\\
{\rm 2\,300\leq Re}\leq4000 & \qquad transitional\\
{\rm Re}>4000 & \qquad turbulent
\end{cases}
\ee
It is clear from the plots in Fig.\ref{fig:RevsT} and Fig.\ref{fig:RevsTamm} that here the flow is always turbulent. In the following we list however the correlation formulas for the three cases, and in Fig.\ref{fig:algo} we provide with a computer algorithm to compute the Nusselt number in any possible configuration.

In the following, we describe the procedure and list the equations which are used to obtain Fig.\ref{fig:algo}. For the formulas we adopt the reference \cite{IncdeVitt}, unless otherwise mentioned.

\subsubsection{Laminar flow}
This corresponds to $\rm Re <2\,300$.
The thermal entry length, defined as
\be
L_{t,laminar}\approx0.05{\rm Re}{\rm Pr}D\,,
\ee
determines if the flow is fully developed, under the condition $L_{t,laminar}<0.5\times L$, where $L=$actual length of the plate. This is practically always the case \cite{IncdeVitt}.
The expression of $\rm Nu$ then depends on the boundary conditions. For a constant surface temperature,
\be
{\rm Nu}=3.66 \qquad T_{s}=constant,
\ee
and for a constant surface heat flux,
\be
{\rm Nu}=4.36\qquad \dot{q}_{x}=constant.
\ee
If instead the flow is not fully developed, we use the following expression of $\rm Nu$ for the entry region of a laminar flow,
\be
{\rm Nu=}3.66+\frac{0.065\left(D/L\right){\rm Re}{\rm Pr}}{1+0.04\left[\left(D/L\right){\rm Re}{\rm Pr}\right]^{2/3}}.
\ee

\subsubsection{Transitional flow}
In this case $2\,300\leq \rm Re\leq4000$.
A transitional flow has a peculiar behavior, hence we can consider it either laminar or turbulent. The following expression is valid for the above range of $\rm Re$,
\be
{\rm Nu=}\frac{\left(f/8\right)\left({\rm Re}-1000\right){\rm Pr}}{1+12.7\left(f/8\right)^{0.5}\left({\rm Pr}^{2/3}-1\right)}\qquad\begin{cases}
0.5\leq{\rm Pr}\leq2000\\
3\times10^{3}<{\rm Re}<5\times10^{6}
\end{cases}
\ee
For other ranges of Prandtl and Reynolds numbers, we consider the flow as turbulent and use the expressions given below in equations (\ref{eq:turb1}), (\ref{eq:turb2}), (\ref{eq:turb3}) and (\ref{eq:turb4}).

\subsubsection{Turbulent flow}
Here $\rm Re >4000$.
The flow is also considered as fully developed, as the thermal entry length is written as
\be
L_{t,turbulent}\approx10D,
\ee
and the pipe diameter is usually rather small as compared to its total length ($L_{t,turbulent}\ll  0.5\times L$). Assume now that the pipe is smooth. The Nusselt number is then given by the Dittus-Boulter equation for cooling,
\be
{\rm Nu}=0.023{\rm Re^{0.8}{\rm Pr}^{0.3}},
\label{eq:turb1}
\ee
which is valid for $0.7\leq {\rm Pr} \leq 160$ and $\rm Re>10000$.

\begin{figure}[t]
\centering
\includegraphics[width=\textwidth]{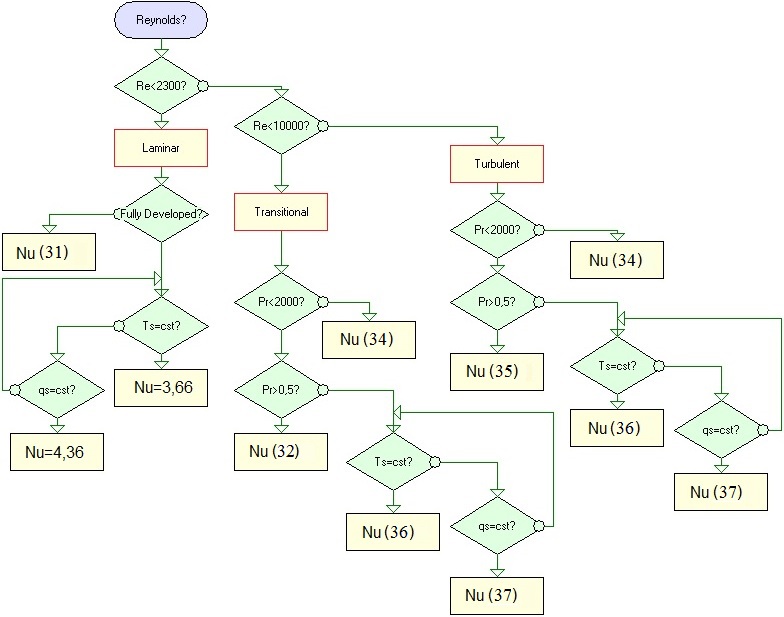}
\caption{Algorithm to find the Nusselt number \cite{Airprop}}
\label{fig:algo}
\end{figure}

For a smaller range of Prandtl numbers, one can use the expression
\be
{\rm Nu}\frac{\left(f/8\right){\rm Re}{\rm Pr}}{1.07+12.7\left(f/8\right)^{0.5}\left({\rm Pr}^{2/3}-1\right)}\qquad\begin{cases}
0.5\leq{\rm Pr}\leq2000\\
10^{4}<{\rm Re}<5\times10^{6}
\end{cases}
\label{eq:turb2}
\ee
which is more accurate. For really small Prandtl numbers the following formulas are available for $10^4\leq \rm Re\leq 10^6$
\be
{\rm Nu}=4.8+0.0156{\rm Re}^{0.85}{\rm Pr}^{0.93}\qquad
{\rm Pr<0.5}, \quad T_{s}=constant,
\label{eq:turb3}
\ee
\be
{\rm Nu}=6.3+0.0167{\rm Re}^{0.85}{\rm Pr}^{0.93}\qquad
{\rm Pr<0.5}, \quad \dot{q}_{s}=constant,
\label{eq:turb4}
\ee
but again it depends on the boundary conditions.
The above considerations are summarized in Fig \ref{fig:algo}, where we suggest an algorithm to find the appropriate Nusselt number \cite{Airprop}.

\begin{figure}[ht]
\begin{center}
\includegraphics[width=0.75\textwidth]{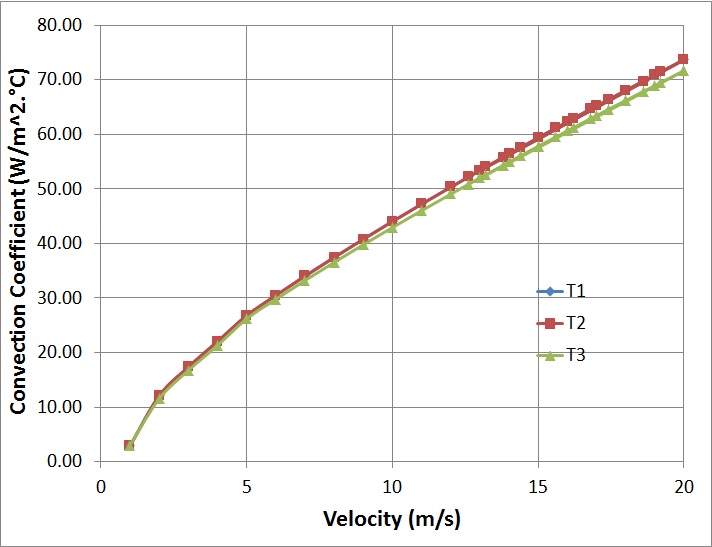}
\caption{$h$ for air in function of the velocity for $D=\unit[20]{cm}$,
$T_{1}=\unit[10]{^\circ C}$, $T_{2}=\unit[15]{^\circ C}$, $T_{3}=\unit[45]{^\circ C}$.
}
\label{fig:hVinfAirInside}
%
\vspace{3mm}

\includegraphics[width=0.75\textwidth]{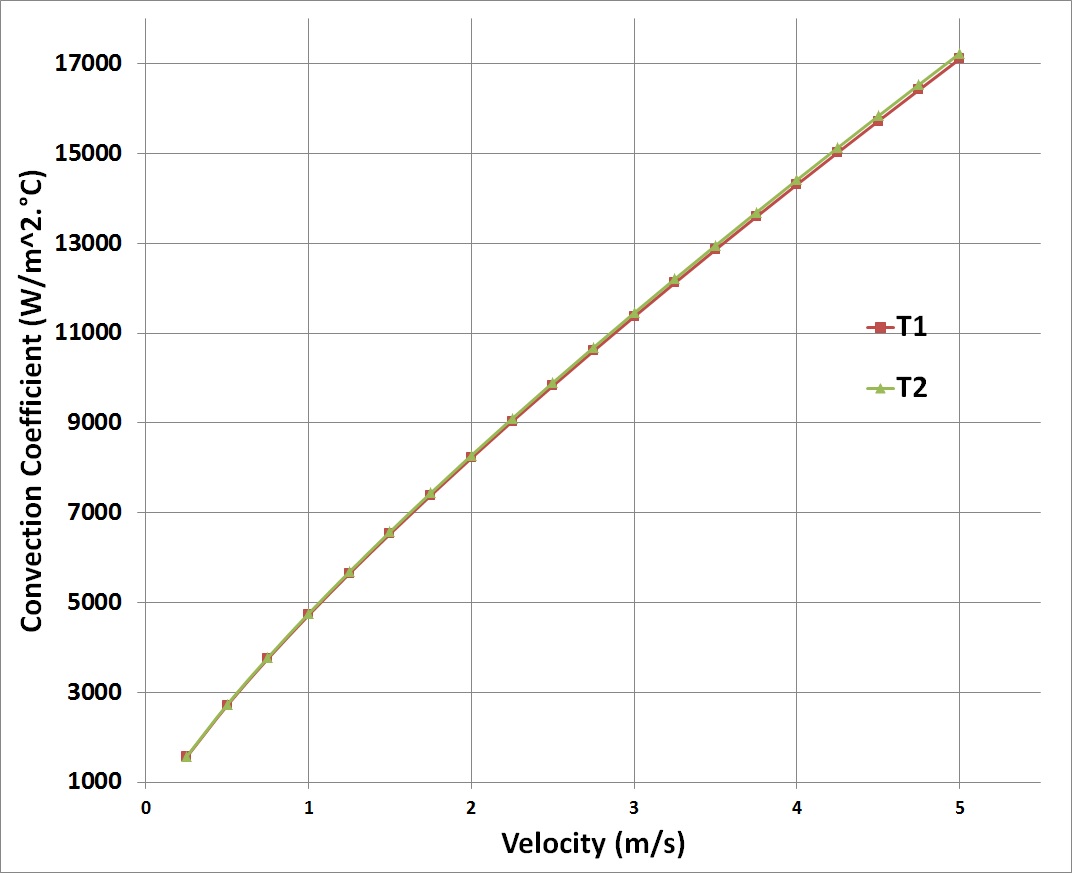}
\caption{$h$ for ammonia in function of the velocity for $D=\unit[3]{cm}$,
 $T_1=\unit[-15]{^\circ C}$, $T_{2}=\unit[-10]{^\circ C}$.}
\label{fig:hVinfRef}
\end{center}
\end{figure}

\begin{figure}[ht]
\begin{center}
\includegraphics[width=0.8\textwidth]{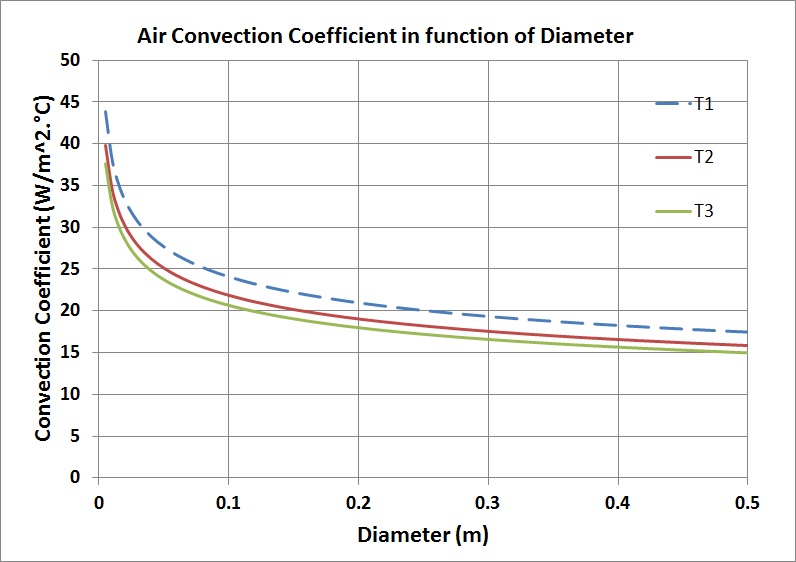}
\caption{Air convection coefficient in function of diameter for $V_\infty=\unit[5]{m/s}$,
$T_{1}=\unit[-30]{^\circ C}$, $T_{2}=\unit[15]{^\circ C}$, $T_{3}=\unit[45]{^\circ C}$.}
\label{fig:hDAir2Inside}
\vspace{3mm}
\includegraphics[width=0.8\textwidth]{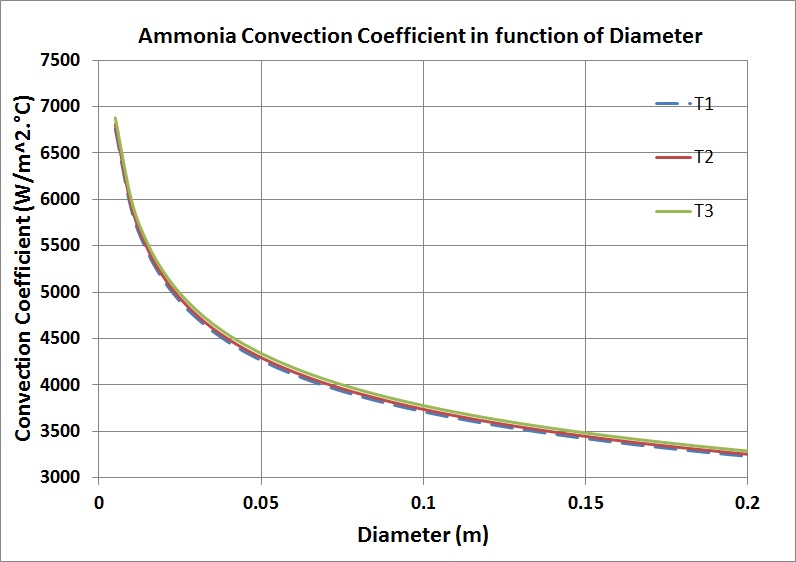}
\caption{Ammonia convection coefficient in function of diameter for $V_\infty=\unit[1]{m/s}$,
$T_{1}=\unit[-15]{^\circ C}$, $T_{2}=\unit[-10]{^\circ C}$, $T_{3}=\unit[-5]{^\circ C}$.}
\label{fig:hDRef2}
\end{center}
\end{figure}

	Table \ref{table:ReynTAirInside} and Table \ref{table:ReynTRef} show that for the case at hand, ${\rm Re}>10000$ and ${\rm Pr}>0.7$, thus we can use Eq.(\ref{eq:turb1}).
Using the common expression Eq.(\ref{eq:hNu}),
one can find the convection coefficient $h$. Figs \ref{fig:hVinfAirInside}, \ref{fig:hVinfRef}, \ref{fig:hDAir2Inside} and \ref{fig:hDRef2} show the dependence of $h$ in function of $V_\infty$ and $D$ at different temperatures \cite{Ash}. This coefficient is proportional to the fluid velocity, as shown in Eq.(\ref{htheo}).

	On the other hand, $h$ decreases by power law when $D$ increases. Especially for air, there is no sensible loss in the heat transfer efficiency for $D>0.2\,m$, see Fig \ref{fig:hDAir2Inside}.

\begin{figure}[ht]
\begin{center}
\includegraphics[width=0.75\textwidth]{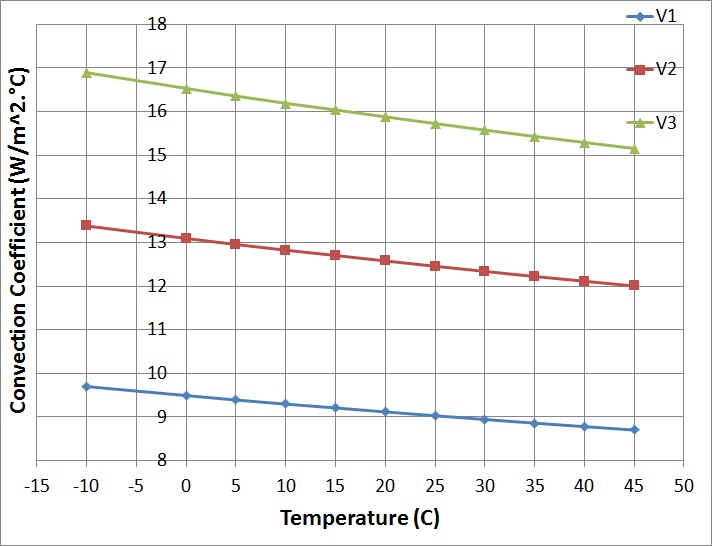}
\caption{$h$ for air in function of the bulk temperature for $D=20\,cm$, $V_{\infty 1}=\unit[2]{m/s}$, $V_{\infty 2}=\unit[3]{m/s}$ and $V_{\infty 3}=\unit[4]{m/s}$.}
\label{fig:hTAir2}
\vspace{3mm}
\includegraphics[width=0.75\textwidth]{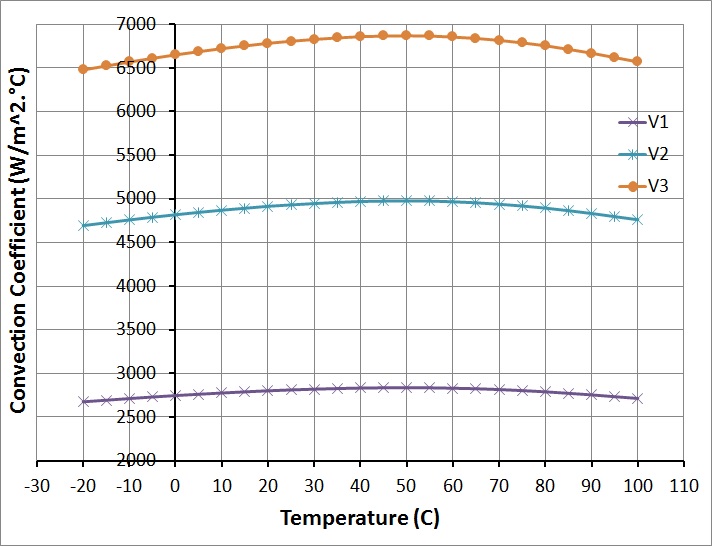}
\caption{$h$ for ammonia in function of the bulk temperature for $D=3\,cm$, $V_{\infty 1}=\unit[0.5]{m/s}$, $V_{\infty 2}=\unit[1]{m/s}$ and $V_{\infty 3}=\unit[1.5]{m/s}$.}
\label{fig:hTRef}
\end{center}
\end{figure}

	One can also see that the bulk temperature has almost no influence on $h$, independently of the temperature range, as in Figs.\ref{fig:hTAir2} and \ref{fig:hTRef}. Notice however the presence of a maximum for ammonia, at $T\sim\unit[50]{^\circ C}$, and the high values for R-717, which justify its use as a refrigerant.

It is useful to derive a precise expression of $h$ in function of the different parameters. The interpolation curves for $h$ hold as follows:
for dry air at 1 atm, $D=\unit[20]{cm}$ and $T=\unit[15]{^\circ C}$,
\be\label{hairV}
h_{air}(V_{\infty})=5.2581\times V_{\infty}^{0.8},
\ee
and for $V_{\infty}=\unit[5]{m/s}$, $T_e=\unit[20]{^\circ C}$, $T=\unit[15]{^\circ C}$,
\be
h_{air}(D)=13.796\times D^{-0.2}.
\ee
For saturated ammonia (R-717), $D=\unit[3]{cm}$ and $T=\unit[-10]{^\circ C}$, we obtain instead
\be\label{hrefV}
h_{ref}(V_{\infty})=4754.3\times V_{\infty}^{0.8},
\ee
and for $V_{\infty}=\unit[1]{m/s}$ and $T=\unit[-5]{^\circ C}$, the equation is
\be
h_{ref}(D)=2384.4\times D^{-0.2}\,.
\ee
The according values of the heat transfer coefficient are plotted in Figs \ref{fig:hVinfAirInside}, \ref{fig:hVinfRef}, \ref{fig:hDAir2Inside} and \ref{fig:hDRef2}, and agree with the literature \cite{Ash}.

It is also relevant to express $h$ as an irreducible function of independent variables, as it was done in Section 2. Let us consider a turbulent flow of a fluid in a duct.
	Since ${\rm Re=}V_{\infty}D/\nu$ and ${\rm Pr}=\mu C_{p}/k$, equation (\ref{eq:turb1}) gives
\be
{\rm Nu}=0.023 \frac{D^{0.8}}{\sqrt{\nu}}\left(\frac{\rho C_p}{k}\right)^{0.3}V_\infty^{0.8},
\ee
therefore using Eq (\ref{eq:hNu}) we obtain
\be\label{htheo}
h(T,D,V_\infty)=0.023\frac{k^{0.7}}{D^{0.2}}\frac{(\rho C_p)^{0.3}}{\sqrt{\nu}}V_\infty^{0.8},
\ee
where $\nu(T)$ and $k(T)$ follow from (\ref{Airnu}) and (\ref{Airk}), and $\rho$ and $C_p$ can be written as functions of temperature as well. The according plots for air and ammonia are given in Figures \ref{fig:hTAir2} and \ref{fig:hTRef}. The interpolation curves are respectively, for air flowing with $V_\infty=2\,m/s$ and $D=20\,cm$,
\be
h(T)=10^{-4}\,T^2 - 0.0191\,T + 9.4151\qquad\left[ \frac{W}{m^2C} \right],
\ee
and for ammonia with $V_\infty=1\,m/s$ and $D=3\,cm$,
\be
h(T)=-0.0679T^2 + 6.2396T + 4826.4\qquad\left[ \frac{W}{m^2C} \right].
\ee
Therefore we find a minimum for air at $T=\unit[75.5]{^\circ C}$, and a maximum for ammonia at $T=\unit[46]{^\circ C}$. Notice that the curve for ammonia has a concavity proportional to the mean bulk velocity.

\clearpage

\section*{Tables and Plots}


\begin{table}[ht]
\begin{minipage}[b]{0.5\linewidth}\centering
\begin{tabular}{c|c|c|c|}
\cline{2-4}
 & $V_{\infty 1}$ & $V_{\infty 2}$ & $V_{\infty 3}$ \\ \cline{2-4}
 & \unit[5]{m/s} & \unit[10]{m/s} & \unit[15]{m/s} \\ \hline
\multicolumn{1}{|c|}{$T(\unit{^\circ C})$} & \multicolumn{3}{|c|}{Reynolds Number ($\times 10^4$)} \\ \hline
\multicolumn{1}{|c|}{-30} & 9.20 & 18.40 & 27.60 \\ \hline
\multicolumn{1}{|c|}{-20} & 8.55 & 17.11 & 25.66 \\ \hline
\multicolumn{1}{|c|}{-10} & 7.99 & 15.97 & 23.96 \\ \hline
\multicolumn{1}{|c|}{0} & 7.47 & 14.95 & 22.42 \\ \hline
\multicolumn{1}{|c|}{10} & 7.01 & 14.03 & 21.04 \\ \hline
\multicolumn{1}{|c|}{20} & 6.60 & 13.19 & 19.79 \\ \hline
\multicolumn{1}{|c|}{30} & 6.22 & 12.44 & 18.66 \\ \hline
\multicolumn{1}{|c|}{40} & 5.88 & 11.75 & 17.63 \\ \hline
\multicolumn{1}{|c|}{50} & 5.56 & 11.12 & 16.69 \\ \hline
\multicolumn{1}{|c|}{60} & 5.27 & 10.55 & 15.82 \\ \hline
\end{tabular}
\caption{Air Reynolds numbers as ${\rm Re } (T,V_{\infty})$, for $D=\unit[20]{cm}$.}
\label{table:ReynTAirInside}
\end{minipage}
\hspace{0.7cm}
\begin{minipage}[b]{0.50\linewidth}
\centering
\begin{tabular}{c|c|c|c|}
\cline{2-4}
 & $V_{\infty 1}$ & $V_{\infty 2}$ & $V_{\infty 3}$ \\ \cline{2-4}
 & \unit[0.75]{m/s} & \unit[2.5]{m/s} & \unit[5]{m/s} \\ \hline
\multicolumn{1}{|c|}{$T(\unit{^\circ C})$} & \multicolumn{3}{|c|}{Reynolds Number ($\times 10^4$)} \\ \hline
\multicolumn{1}{|c|}{-40} & 4.78 & 15.92 & 31.84 \\ \hline
\multicolumn{1}{|c|}{-30} & 5.22 & 17.40 & 34.79 \\ \hline
\multicolumn{1}{|c|}{-25} & 5.46 & 18.19 & 36.38 \\ \hline
\multicolumn{1}{|c|}{-20} & 5.70 & 19.01 & 38.03 \\ \hline
\multicolumn{1}{|c|}{-15} & 5.96 & 19.88 & 39.76 \\ \hline
\multicolumn{1}{|c|}{-10} & 6.24 & 20.79 & 41.58 \\ \hline
\multicolumn{1}{|c|}{-5} & 6.52 & 21.75 & 43.50 \\ \hline
\multicolumn{1}{|c|}{0} & 6.82 & 22.74 & 45.47 \\ \hline
\end{tabular}
\caption{Ammonia Reynolds numbers as ${\rm Re } (T,V_{\infty})$,  for $D=\unit[27]{mm}$
.}
\label{table:ReynTRef}
\end{minipage}
\end{table}


\begin{table}[ht]
\begin{minipage}[b]{0.45\linewidth}\centering
\begin{tabular}{c|c|c|c|}
\cline{2-4}
 & $T_1$ & $T_2$ & $T_3$ \\ \cline{2-4}
 & \unit[10]{$^\circ$ C} & \unit[15]{$^\circ$ C} & \unit[45]{$^\circ$ C} \\ \hline
\multicolumn{1}{|c|}{$V_{\infty}$} & \multicolumn{3}{|c|}{Reynolds Number ($\times 10^4$)} \\ \hline
\multicolumn{1}{|c|}{1} & 1.40 & 1.36 &	1.14 \\ \hline
\multicolumn{1}{|c|}{2} & 2.80 & 2.72 & 2.29 \\ \hline
\multicolumn{1}{|c|}{3} & 4.20 & 4.08 & 3.43 \\ \hline
\multicolumn{1}{|c|}{4} & 5.59 & 5.44 & 4.57 \\ \hline
\multicolumn{1}{|c|}{5} & 6.99 & 6.80 & 5.71 \\ \hline
\multicolumn{1}{|c|}{6} & 8.39 & 8.16 & 6.86 \\ \hline
\multicolumn{1}{|c|}{7} & 9.79 & 9.52 & 8.00 \\ \hline
\multicolumn{1}{|c|}{8} & 11.19 & 10.88 & 9.14 \\ \hline
\multicolumn{1}{|c|}{9} & 12.59 & 12.25 & 10.29 \\ \hline
\multicolumn{1}{|c|}{10} & 13.99 & 13.61 & 11.43 \\ \hline
\multicolumn{1}{|c|}{11} & 15.38 & 14.97 & 12.57 \\ \hline
\multicolumn{1}{|c|}{12} & 16.78 & 16.33 & 13.71 \\ \hline
\multicolumn{1}{|c|}{13} & 18.18 & 17.69 & 14.86 \\ \hline
\multicolumn{1}{|c|}{14} & 19.58 & 19.05 & 16.00 \\ \hline
\multicolumn{1}{|c|}{15} & 20.98 & 20.41 & 17.14 \\ \hline
\end{tabular}
\caption{${\rm Re}(V_{\infty},T)$ for air, $D=\unit[20]{cm}$ and $[V_\infty]=[m/s]$.}
\label{table:ReynVAirInside}
\end{minipage}
\hspace{0.2cm}
\begin{minipage}[b]{0.5\linewidth}
\centering
\begin{tabular}{c|c|c|}
\cline{2-3}
 & $T_1$ & $T_2$ \\ \cline{2-3}
 & \unit[-15]{$^\circ$ C} & \unit[-9]{$^\circ$ C}\\ \hline
\multicolumn{1}{|c|}{$V_{\infty}$} & \multicolumn{2}{|c|}{Reynolds Number ($\times 10^4$)} \\ \hline
\multicolumn{1}{|c|}{0.25} & 1.99 & 2.08 \\ \hline
\multicolumn{1}{|c|}{0.5} & 3.98 & 4.16 \\ \hline
\multicolumn{1}{|c|}{0.75} & 5.97 & 6.24 \\ \hline
\multicolumn{1}{|c|}{1} & 7.95 & 8.32  \\ \hline
\multicolumn{1}{|c|}{1.25} & 9.94 & 10.40 \\ \hline
\multicolumn{1}{|c|}{1.5} & 11.93 & 12.48 \\ \hline
\multicolumn{1}{|c|}{1.75} & 13.92 & 14.55 \\ \hline
\multicolumn{1}{|c|}{2} & 15.91 & 16.63  \\ \hline
\multicolumn{1}{|c|}{2.25} & 17.89 & 18.71 \\ \hline
\multicolumn{1}{|c|}{2.5} & 19.88 & 20.79 \\ \hline
\multicolumn{1}{|c|}{2.75} & 21.87 & 22.87 \\ \hline
\multicolumn{1}{|c|}{3} & 23.86 & 24.95  \\ \hline
\multicolumn{1}{|c|}{3.25} & 25.85 & 27.03 \\ \hline
\multicolumn{1}{|c|}{3.5} & 27.83 & 29.11  \\ \hline
\multicolumn{1}{|c|}{3.75} & 29.82 & 31.19  \\ \hline
\multicolumn{1}{|c|}{4} & 31.81 & 33.27 \\ \hline
\multicolumn{1}{|c|}{4.25} & 33.80 & 35.35 \\ \hline
\multicolumn{1}{|c|}{4.5} & 35.79 & 37.43 \\ \hline
\multicolumn{1}{|c|}{4.75} & 37.78 & 39.50 \\ \hline
\multicolumn{1}{|c|}{5} & 39.76 & 41.58 \\ \hline\end{tabular}
\caption{${\rm Re}(V_{\infty},T)$ for ammonia, $D=\unit[27]{mm}$ and $[V_\infty]=[m/s]$.}
\label{table:ReynVRef}
\end{minipage}
\end{table}

\begin{table}[ht]
\begin{minipage}[b]{0.50\linewidth}
\begin{center}\begin{tabular}{c|c|c|c|}
\cline{2-4}
 & $T_{1}$ & $T_2$ & $T_3$ \\ \cline{2-4}
 & \unit[10]{$^\circ$ C} & \unit[15]{$^\circ$ C} & \unit[45]{$^\circ$ C} \\ \hline
\multicolumn{1}{|c|}{$V_{\infty}$(\unit{m/s})} & \multicolumn{3}{|c|}{$h\unit(W/m^{2}. ^{\circ} C)$} \\ \hline
\multicolumn{1}{|c|}{1} & 5.30 & 5.26 & 4.97 \\ \hline
\multicolumn{1}{|c|}{2} & 9.22 & 9.15 & 8.65 \\ \hline
\multicolumn{1}{|c|}{3} & 12.76 & 12.66 & 11.97 \\ \hline
\multicolumn{1}{|c|}{4} & 16.06 & 15.94 & 15.06 \\ \hline
\multicolumn{1}{|c|}{5} & 19.20 & 19.05 & 18.01 \\ \hline
\multicolumn{1}{|c|}{6} & 22.21 & 22.05 & 20.83 \\ \hline
\multicolumn{1}{|c|}{7} & 25.13 & 24.94 & 23.57 \\ \hline
\multicolumn{1}{|c|}{8} & 27.96 & 27.75 & 26.22 \\ \hline
\multicolumn{1}{|c|}{9} & 30.73 & 30.49 & 28.82 \\ \hline
\multicolumn{1}{|c|}{10} & 33.43 & 33.18 & 31.35 \\ \hline
\multicolumn{1}{|c|}{11} & 36.08 & 35.80 & 33.83 \\ \hline
\multicolumn{1}{|c|}{12} & 38.68 & 38.39 & 36.27 \\ \hline
\multicolumn{1}{|c|}{13} & 41.23 & 40.92 & 38.67 \\ \hline
\multicolumn{1}{|c|}{14} & 43.75 & 43.42 & 41.03 \\ \hline
\multicolumn{1}{|c|}{15} & 46.24 & 45.89 & 43.36 \\ \hline
\multicolumn{1}{|c|}{16} & 48.69 & 48.32 & 45.66 \\ \hline
\multicolumn{1}{|c|}{17} & 51.10 & 50.72 & 47.93 \\ \hline
\multicolumn{1}{|c|}{18} & 53.50 & 53.09 & 50.17 \\ \hline
\multicolumn{1}{|c|}{19} & 55.86 & 55.44 & 52.39 \\ \hline
\multicolumn{1}{|c|}{20} & 58.20 & 57.76 & 54.58 \\ \hline
\end{tabular}
\caption{Air convection coefficient in function of the velocity and $T$ for $D=\unit[20]{cm}$
}
\label{table:hVinfVAir}
\end{center}
\end{minipage}
\hspace{0.2cm}
\begin{minipage}[b]{0.45\linewidth}
\begin{center}\begin{tabular}{c|c|c|}
\cline{2-3}
 & $T_1$ & $T_2$  \\ \cline{2-3}
 & \unit[-15]{$^\circ$ C}  & \unit[-10]{$^\circ$ C} \\ \hline
\multicolumn{1}{|c|}{$V_{\infty}$(\unit{m/s})} & \multicolumn{2}{|c|}{$h\unit(W/m^{2}. ^{\circ} C)$} \\ \hline
\multicolumn{1}{|c|}{0.25} & 1557  & 1568  \\ \hline
\multicolumn{1}{|c|}{0.5} & 2711  & 2730 \\ \hline
\multicolumn{1}{|c|}{0.75} & 3750 & 3777 \\ \hline
\multicolumn{1}{|c|}{1} & 4720  & 4754 \\ \hline
\multicolumn{1}{|c|}{1.25} & 5643  & 5684 \\ \hline
\multicolumn{1}{|c|}{1.5} & 6529  & 6576 \\ \hline
\multicolumn{1}{|c|}{1.75} & 7386  & 7439 \\ \hline
\multicolumn{1}{|c|}{2} & 8219  & 8278 \\ \hline
\multicolumn{1}{|c|}{2.25} & 9031  & 9096 \\ \hline
\multicolumn{1}{|c|}{2.5} & 9825  & 9896 \\ \hline
\multicolumn{1}{|c|}{2.75} & 10604  & 10680 \\ \hline
\multicolumn{1}{|c|}{3} & 11368  & 11449 \\ \hline
\multicolumn{1}{|c|}{3.25} & 12120  & 12207 \\ \hline
\multicolumn{1}{|c|}{3.5} & 12860  & 12952 \\ \hline
\multicolumn{1}{|c|}{3.75} & 13590  & 13687 \\ \hline
\multicolumn{1}{|c|}{4} & 14310 &  14412 \\ \hline
\multicolumn{1}{|c|}{4.25} & 15021 &  15129 \\ \hline
\multicolumn{1}{|c|}{4.5} & 15724 &  15836 \\ \hline
\multicolumn{1}{|c|}{4.75} & 16419 &  16536 \\ \hline
\multicolumn{1}{|c|}{5} & 17106 &  17229 \\ \hline
\end{tabular}
\caption{Ammonia convection coefficient in function of the velocity and $T$ for $D=\unit[3]{cm}$.
}
\label{table:hVinfVRef}
\end{center}
\end{minipage}
\end{table}


%
%
%
%
%
%
%


\clearpage

\appendix
\section{Critical length and least square method}\label{appendix}

The least square method (see e.g. \cite{Leastsquare}) is widely used in several fields, for instance in experimental physics, and dates back to the first works by Carl Friedrich Gauss and Legendre. In this section we apply this procedure to write $\nu$ as a function of temperature in two cases of fluid flow over a hot plate, namely unused engine oil and dry air at 1 atm. We also compute the critical length in function of the fluid temperature.


According to experimental data (\cite{IncdeVitt}), $\nu$ has the shape of a power function, which can be written as
\be
\nu=kT^{-m},
\ee
thus one needs to determine $k$ and $m$. We start with an ansatz on these, and then plot the equation with those values. Then we
calculate the difference between the empirical $\nu$ and the one obtained with the formula. Then we compute the square sum of
these differences and minize this vector by changing the two parameters $k$ and $m$. This is done automatically with several programs, giving
\be
\nu(T)=6.1875\times T^{-2.7427}.
\ee
Note that this function is valid only for the temperature of use of the engine oil (above $40^{\circ}$C), otherwise the approximation is a lot less precise. Here the average error is 0,9\%.

\begin{figure}[ht]
\centering
\includegraphics[width=\textwidth]{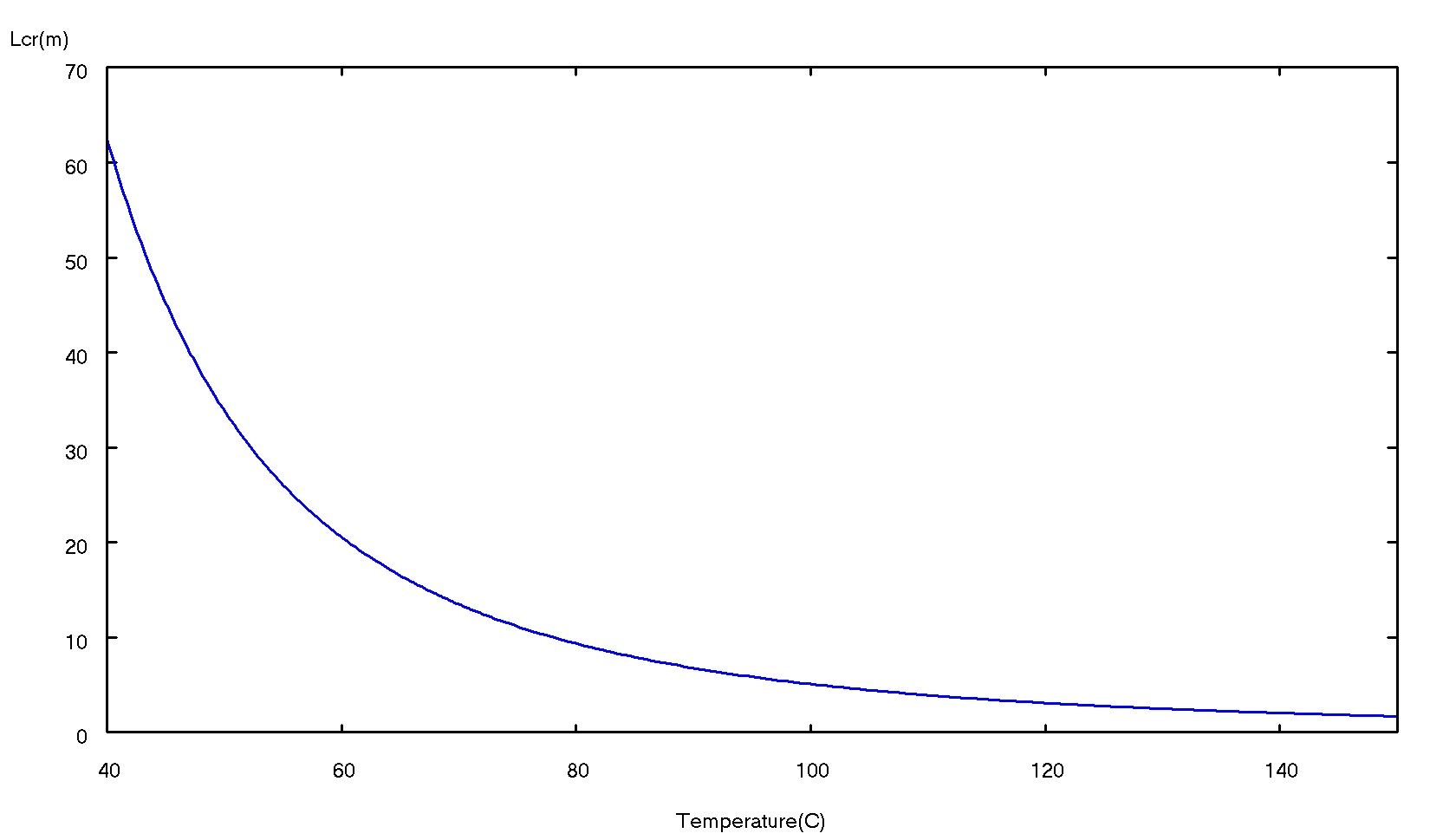}
\caption{Oil Critical Length in function of temperature.}
\label{LcrOil}

\includegraphics[width=\textwidth]{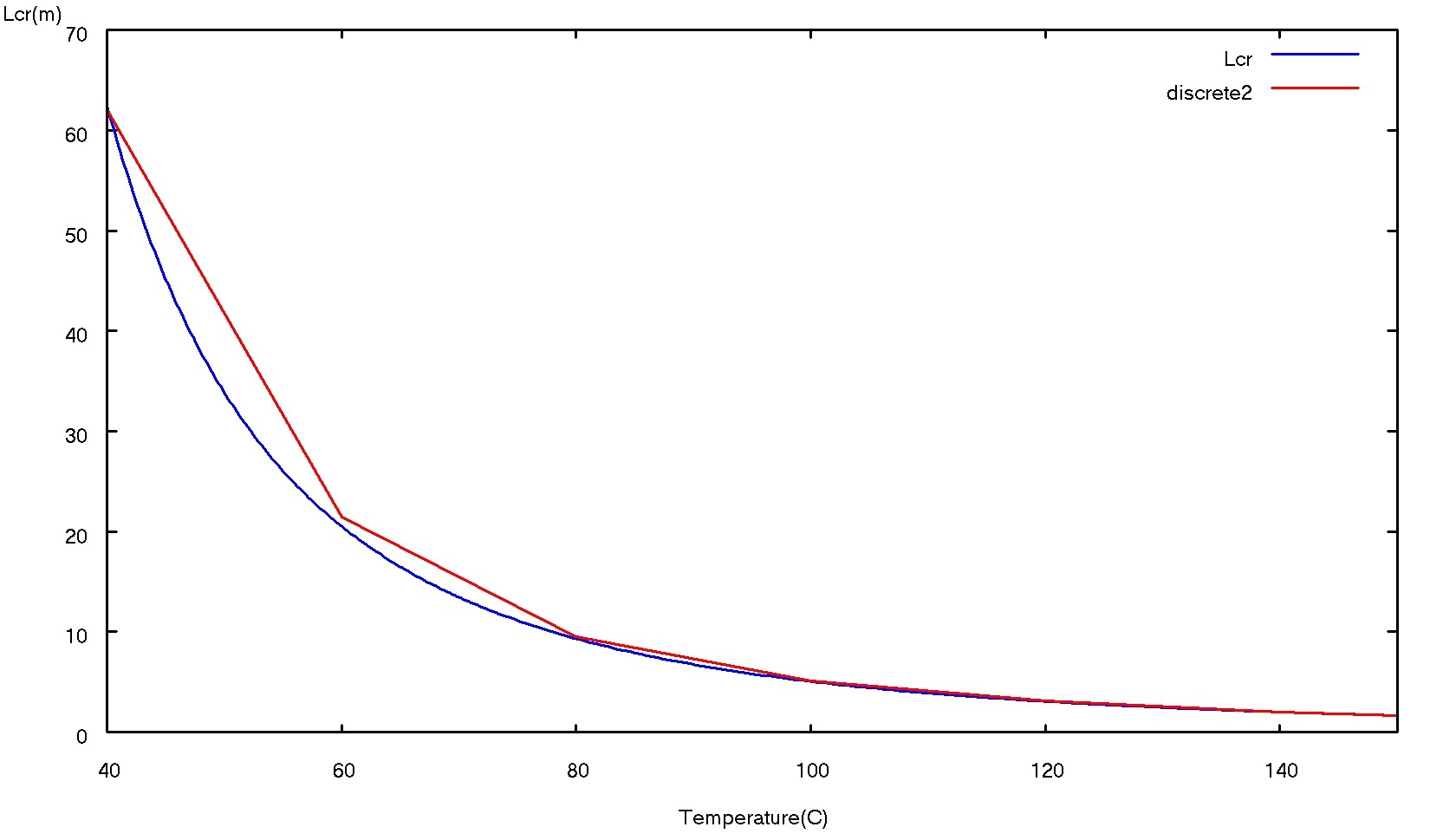}
\caption{Oil Critical Length in function of temperature: comparison with empirical points.}
\label{LcrExpOil}
\end{figure}

Next, one can express $L_{cr}$ as an irreducible function of T,
\be
L_{cr}(T)=\frac{\nu(T)}{V_{\infty}}Re_{cr}=\frac{6.1875\times T^{-2.7427}}{V_{\infty}}5\times10^{5}.
\ee
Fig \ref{LcrOil} gives the precise behavior of $L_{cr}$ in function of the oil temperature. Here we set $V_{\infty}=2m/s$ for the
oil.

In Fig \ref{LcrExpOil} we see the juxtaposition of the two curves (empirical and analytical). However between $40^{\circ}C$ and $60^{\circ}C$ the analytical curve does not fit exactly the experimental points. This gap comes from the exponential behavior of the function. 

\begin{figure}[ht]
\centering
\includegraphics[width=\textwidth]{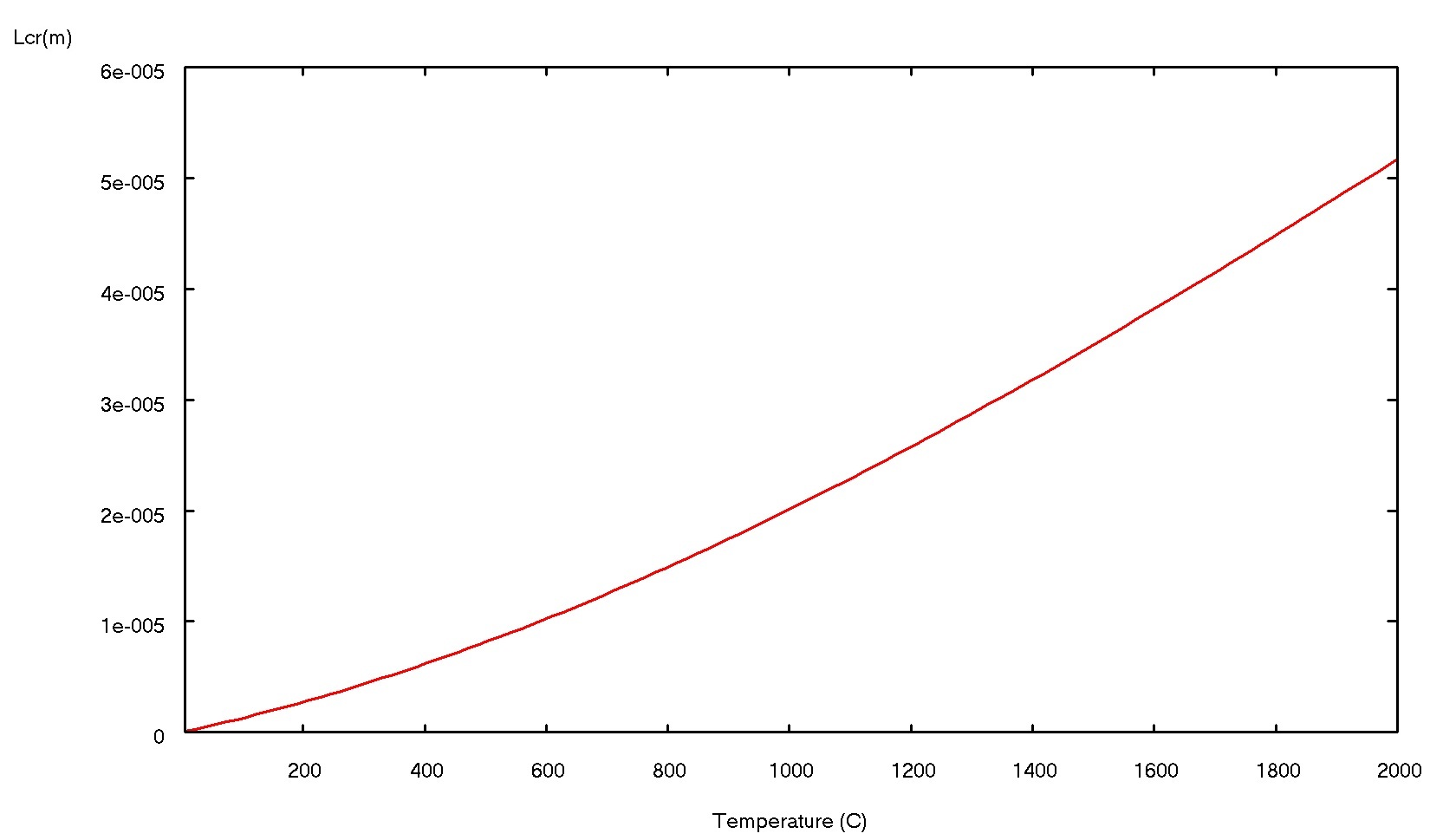}
\caption{Air Critical Length in function of temperature.}
\label{LcrAir}

\includegraphics[width=\textwidth]{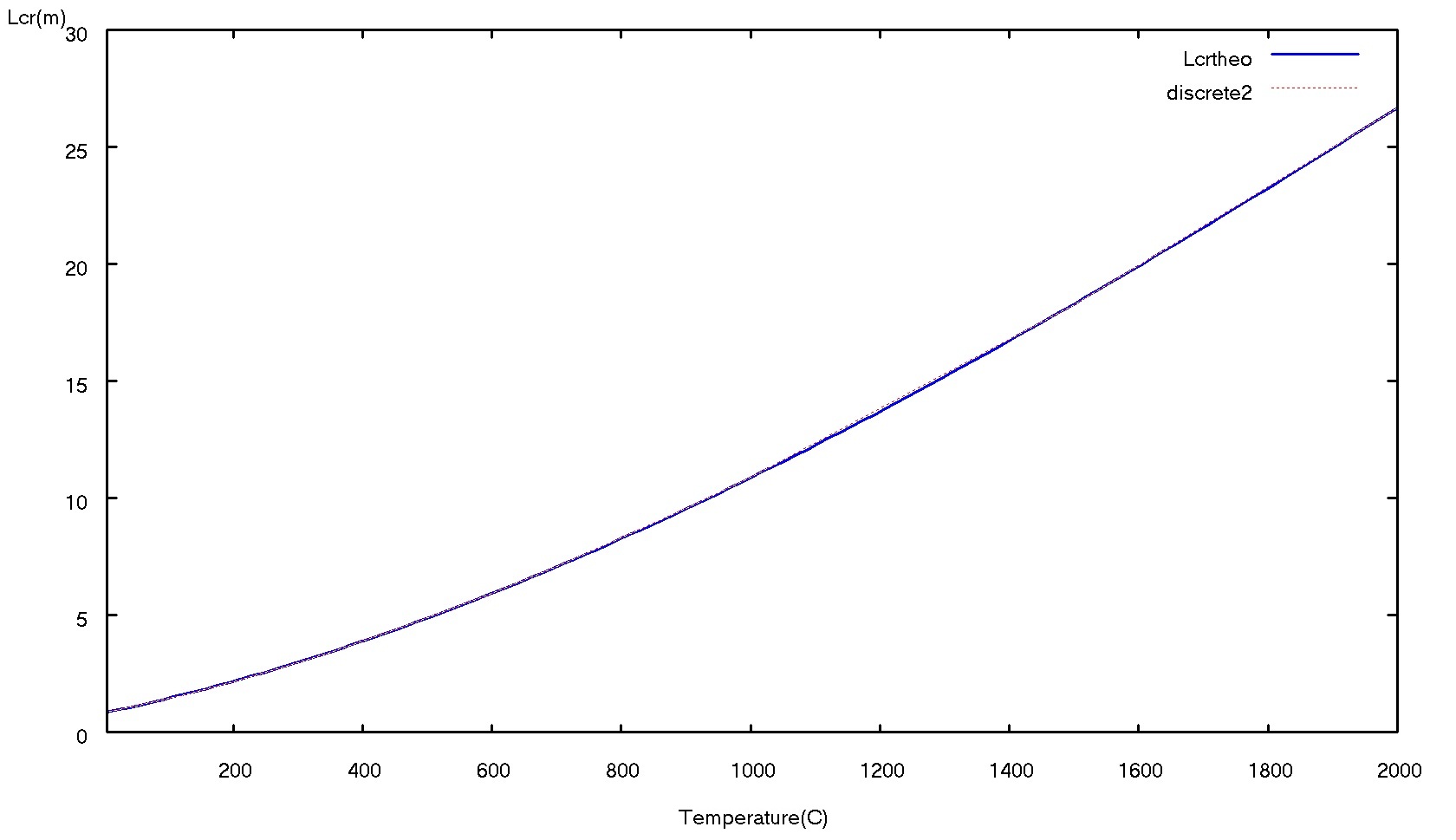}
\caption{Air Critical Length in function of temperature: comparison with empirical points.}
\label{LcrExpAir}
\end{figure}

Consider now dry air at 1 atm. THe kinematic viscosity $\nu$ has the shape of a polynomial function: $\nu=aT^{2}+bT+c$, thus
we want to find $a$, $b$ and $c$.
Using the same solver method than with oil we find:
\be
\nu\left(T\right)=4.91452\times10^{-11}T^{2}+1.10682\times10^{-7}T+1.18052\times10^{-5},
\ee
with units $[m^2/s]$.
Likewise, the above formula is valid only for the temperature of use (above 5$^{\circ}$C). Here we have an average error of 0.98\%. Note that a third degree polynomial function is fairly more accurate,
\be
\nu\left(T\right)=-1.0743\times10^{-14}T^{3}+7.8253\times10^{-11}T^{2}+9.33152\times10^{-8}T+1.3094\times10^{-5},
\ee
with an average error of 0.09\%.
	Therefore we can express $L_{cr}$ as a function of T as follows,
\bea
&&L_{cr}(T)=\frac{\nu(T)}{V_{\infty}}Re_{cr}
=\frac{5}{V_{\infty}}\times
\Big(
\nonumber\\
&& 
-1.0743\times10^{-9}\,T^{3}+7.8253\times10^{-6}\,T^{2}+9.33152\times10^{-3}\,T+1.3094
\Big).\nonumber\\
\eea
Here we set $V_{\infty}=8m/s$ for air, see Fig \ref{LcrAir}. 
	Fig \ref{LcrExpAir} compares the empirical points and the analytical curve. However the third degree polynomial function is so precise that we cannot even discern gaps between the two curves.

\clearpage

\listoftables
\listoffigures

\clearpage

\section*{Conclusions and perspectives}

In this paper we have discussed an method to the study of heat transfer processes, which is aimed at academic teaching but can be useful also in practical situations. It has analogy with the so-called \emph{functional optimization}, see \cite{Opt} and references quoted therein.

This approach focuses on a theoretical analysis of the relevant equations, and makes it possible to know the behavior of the crucial quantities \emph{a priori}, before any empirical tests. This creates awareness of the physics process in the students, and aids researchers effectively.

	We have applied this method to two common heat transfer processes, namely to 1. a fluid flow across a cylinder and 2. a pipe flow. In both cases we have discussed the behavior of the Reynolds number and of the convection coefficient, in function of variables such as fluid temperature and pipe diameter, using realistic test values.
	
	Our results, listed in a number of plots and tables, can be directly used as a reference for the students in academic courses. In the Appendix, we also show how to use the least square method to compute interpolation curves.


\end{document}